\documentclass[12pt,aps,tightenlines,showpacs,showkeys]{revtex4}

\usepackage{amsmath}
\usepackage{amsfonts}
\usepackage{bm}
\usepackage{graphicx}
\usepackage{subfigure}

\graphicspath{{figs_pre/}}

\begin{document}

\title{Bubbles and Filaments: Stirring a Cahn--Hilliard Fluid}
\author{Lennon \'O N\'araigh}
\author{Jean-Luc Thiffeault}
\email{jeanluc@imperial.ac.uk}
\affiliation{Department of Mathematics, Imperial College
  London, SW7 2AZ, United Kingdom}
\date{\today}

\begin{abstract}
        The advective Cahn--Hilliard equation describes the competing
        processes of stirring and separation in a two-phase fluid.  Intuition
        suggests that bubbles will form on a certain scale, and previous
        studies of Cahn--Hilliard dynamics seem to suggest the presence of one
        dominant length scale.  However, the Cahn--Hilliard phase-separation
        mechanism contains a hyperdiffusion term and we show that, by stirring
        the mixture at a sufficiently large amplitude, we excite the diffusion
        and overwhelm the segregation to create a homogeneous liquid.  At
        intermediate amplitudes we see regions of bubbles coexisting with
        regions of hyperdiffusive filaments.  Thus, the problem possesses two
        dominant length scales, associated with the bubbles and filaments.
        For simplicity, we use use a chaotic flow that mimics turbulent
        stirring at large Prandtl number.  We compare our results with the case
        of variable mobility, in which growth of bubble size is
        dominated by interfacial rather than bulk effects, and find
        qualitatively similar results.
\end{abstract}



\maketitle

\section{Introduction}
\label{sec:Introduction}
In a seminal paper, Cahn and Hilliard \cite{CH_orig} introduced their
eponymous equation to model the dynamics of phase separation.  They pictured
a
binary alloy in a mixed state, cooling below a critical temperature.  This
state is unstable to small perturbations so that fluctuations cause the alloy
to separate into bubbles (or domains) rich in one material or the other.  A
small transition layer separates the bubbles.  Because the evolution of the
concentration field is an order parameter equation, this description is
completely general.  Thus, by coupling the phase-separation model to fluid
equations, one has a broad description of binary fluid flow encompassing
polymers, immiscible binary fluids with interfacial tension, and glasses.

In industrial applications, the components of a binary mixture often need to
be mixed in a homogeneous state.  In that case, the coarsening tendency of
multiphase fluids is undesirable.  In this paper we will examine how a
stirring flow affects this mixing process.  We are interested in stirring not
just to limit bubble growth, as is often effected with shear flows, but to
break up the bubbles into a homogeneous mixture.

We first outline previous work for three increasing levels of complexity:
Cahn--Hilliard (CH) in the absence of flow, passive CH with flow, and active
CH with flow.  Our focus in this paper will be on the passive case.

\emph{Cahn--Hilliard fluids without flow:} There is a substantial literature
that treats of the CH equation without flow.  For a review, see
\cite{Bray_advphys}.  The main physical feature of this simpler model is the
existence of a single length scale (the bubble or domain size) such that
quantities of interest (e.g., the correlation function) are self-similar
under this scale.  This length grows in time as $t^{1/3}$, a result
seen in many numerical simulations (see, for example, Zhu et
al.~\cite{Zhu_numerics}).  Of mathematical interest are the existence of a
Lyapunov functional, and hence of a bounded solution \cite{Elliott_Zheng}, and
the failure of the maximum principle owing to fourth-order derivatives in the
equation for the concentration field \cite{Gajewski_nonlocal,Elliott_varmob}.

\emph{Passive Cahn--Hilliard fluids:} Given an externally imposed flow (e.g.,
a shear flow, turbulence, or a stirring motion), we have a passive tracer
equation for a Cahn--Hilliard (CH) fluid.  In the same way that one invokes
the mixing of coffee and cream to picture the mechanisms at work in the
advection-diffusion equation, this is the chef's problem of how to mix olive
oil and soy sauce by stirring.

There have been several studies of the (active and passive) shear-driven CH
fluid \cite{shear_Shou, shear_Berthier}.  In these studies, it is claimed that
domain elongation persists in a direction that asymptotically aligns with the
flow direction, while domain formation is arrested in a direction
perpendicular to this axis.  This seems to be confirmed in the experiments of
Hashimoto~\cite{Hashimoto}, although these results could be due to finite-size
effects~\cite{shear_Bray}, in which the arrest of domain growth is due to
the limitation of the boundary on domain size.   Of
interest is the existence or otherwise of an arrest scale and the dependence
of this scale on the system parameters.

A more effective form of stirring involves chaotic flows, in which neighboring
particle trajectories separate exponentially in time, leading to chaotic
advection~\cite{Aref1984}.  The average rate of separation is the Lyapunov
exponent, positive for such a flow~\cite{Eckmann1985,WigginsIntroDynSys}.  One
may also regard this exponent as the average rate of strain of the flow.  By
imposing a chaotic flow on the CH fluid, finite-size effects are easily
overcome. This is because the shear direction changes in time, so that domain
elongtation in one
particular direction, leading to an eventual steady state in which domains
touch many walls, is no longer possible.  Berthier et
al. \cite{chaos_Berthier} investigated the passive stirring of the CH fluid
and observed a coarsening arrest arising from a dynamical balance of surface
tension and advection.  The stirring by the chaotic flow destroys large-scale
structures, and this is balanced by the domain-forming tendency of the CH
equation.  A balancing scale is identified with the equilibrium domain size.

The importance of exponential stretching is amplified by the results of
Lacasta et al. \cite{Sancho}, in which a passive turbulent flow endowed with a
tunable amplitude, correlation length, and correlation time, causes both
arrest and indefinite growth, in different limits.  We explain this
heuristically as follows: In one limit of turbulent flow (small Prantdl
number), the correlation length of the turbulent velocity field is small
compared to the typical bubble size, fluid particles
diffuse~\cite{Taylor1921}, and bubble radii evolve as ${d R_b^2}/dt =
2\kappa_{\mathrm{eff}} + \text{[Cahn--Hilliard contribution]}$, where $R_b$ is
the bubble radius and $\kappa_{\mathrm{eff}}$ is the effective tracer
diffusivity.  Because $\kappa_{\mathrm{eff}} > 0$, the presence of diffusion
fails to arrest bubble growth.  In the limit of large Prantdl number however,
the the correlation length is large and the velocity field gives rise to
exponential separation of trajectories, so that the radius equation is ${d
R_b^2}/{dt}=-2\lambda R_b^2+ \text{[Cahn--Hilliard contribution]}$, where
$\lambda$ is the average rate of strain, or Lyapunov exponent, of the flow.
Thus, the exponential stretching due to stirring balances the bubble growth.
Lacasta et al. \cite{Sancho} identify a critical correlation length above
which coarsening arrest always takes place, no matter how small the stirring
amplitude.  Since chaotic flows typically exhibit long-range correlations, we
expect they will always produce coarsening arrest.

\emph{Active Cahn--Hilliard fluids:} At the highest level of complexity, one
considers the reaction of the mixture on the flow by coupling the CH equation
to the Navier--Stokes equations.  For a derivation of the Navier--Stokes
Cahn--Hilliard equations, see \cite{LowenTrus}.  In experiments involving
turbulent binary fluids near the critical temperature, it is found that a
turbulent flow suppresses phase separation and homogenizes the fluid
\cite{Pine1984,Chan1987}.  However, by cooling the fluid further and
maintaining the same level of forcing, phase separation is achieved.  This
result is limited to near-critical fluids, although it does suggest the
possibility of homogenization by stirring in other contexts.

The most recent work on the
CH fluid (Berti et al., \cite{Berti_goodstuff}) focusses on the active CH
tracer.  They couple the CH concentration field to an externally forced
velocity field and observe coarsening arrest independently of finite-size
effects.  They identify a balance of local shears and surface tension as the
mechanism for coarsening arrest.  Thus, the chaotic advection discussed above
is a sufficient condition for coarsening arrest, but not a necessary one since
shear will suffice.  Of interest again is the dependence of the arrest scale
on the mean rate of shear.

With this literature in mind, we study the case of a chaotic flow coupled
passively to the CH equation.  Our ultimate focus will be on the breakup of
bubbles by stirring, and not just coarsening arrest.  We believe this aspect
of CH flow deserves a further and more complete study, as it is a regime of
particular relevance to industry, for example, in droplet breakup or in the
mixing of polymers \cite{Aarts}.

Berthier et al.~\cite{chaos_Berthier} used an alternating sine flow to study
coarsening arrest.  Because of its simplicity, the sine flow is a popular
testbed for studying chaotic
mixing~\cite{lattice_PH2,Antonsen1996,Neufeld_filaments,lattice_PH1,%
Thiffeault2004}.  In contrast to Berthier et al.~\cite{chaos_Berthier}, we use
a sine flow where the phase is randomized at each period (see
Section~\ref{sec:numerics}), leading to a flow that has a positive Lyapunov
exponent at all stirring amplitudes and for all initial conditions.  This
avoids the coexistence of regular and chaotic regions, which would lead to
inhomeogeneous mixing characteristics over the domain of the problem.  Because of its uniform
stirring, the random-phase sine flow is often used as a proxy for turbulent
stirring at high Prandtl number, but at much lower computational cost.  Using
this flow, we study coarsening arrest, as in previous papers
\cite{chaos_Berthier, Berti_goodstuff}.  However, we shall also use it to
study mixing due to the stirring and the hyperdiffusion of the CH dynamics.
Under this additional process the scaling hypothesis that characterizes the
dynamics of bubble formation, namely that there is one dominant length scale~\cite{Bray_advphys,
Berti_goodstuff}, breaks down.

Following studies made Batchelor on the advection-diffusion equation~\cite{Batchelor1959}
and by Neufeld on reaction-diffusion equations \cite{Neufeld_filaments},
we introduce a one-dimensional equation in order
to model the formation of bubbles and the influence of advection and hyperdiffusion
on the bubble size.  This will lead to an understanding of the mechanism
by which the bubbles are broken up and hyperdiffusion is becomes dominant.

The paper is laid out in the following way.  In Section~\ref{sec:model} we
introduce the CH equation with advection and discuss its scaling laws.  In
Section~\ref{sec:numerics} we present some numerical simulations of the
two-dimensional equations, with and without flow, and confirm the scaling
laws.  In Section~\ref{sec:1D} we make use of the one-dimensional CH equation
coupled to a straining flow to understand the results from the two-dimensional
simulations.  In Section~\ref{sec:varmob} we investigate numerically the
two-dimensional CH equation with variable mobility.
\section{The Model Equation and its Scaling Laws}
\label{sec:model}
In this section we introduce the advective Cahn--Hilliard equation in $n$
dimensions.  We then discuss the notion of dynamical equilibrium for the CH
equation without flow in terms of the structure function.  Finally, we study
the length scales that arise in the presence of flow.

Let $c\left(\bm{x},t\right)$ be the conserved order parameter of the problem
and let $\bm{v}\left(\bm{x},t\right)$ be an externally imposed two-dimensional
incompressible flow, $\nabla\cdot\bm{v}=0$.  The advective Cahn--Hilliard
equation describes the phase separation dynamics of the field
$c\left(\bm{x},t\right)$ in the presence of flow,
\begin{subequations}
\begin{gather}
  \frac{\partial c}{\partial t} +\bm{v}\cdot\nabla c =
  D\nabla^2\left(f_0'\left(c\right)-\gamma\nabla^2
c\right),\\
  f_0\left(c\right) = \tfrac{1}{4}\left(c^2-1\right)^2.
\end{gather}%
\label{eq:CH}%
\end{subequations}%
This equation characterizes the stirring of a phase-separating binary liquid.
Here $D$ is the (constant) mobility, which governs the rate of phase
separation, and $\gamma$ is a hyperdiffusion coefficient.  The fourth-order
hyperdiffusion term in Eq.~\eqref{eq:CH} has a mollifying effect that prevents
the blow up of gradients.  By identifying a chemical potential $\mu =
f_0'\left(c\right) -\gamma\nabla^2 c$, Eq.~\eqref{eq:CH} is recast as
\begin{equation}
  \frac{\partial c}{\partial t}+\bm{v}\cdot\nabla c = D\nabla^2\mu\,.
\label{eq:CHmu}
\end{equation}
The form of Eqs.~\eqref{eq:CH} and~\eqref{eq:CHmu} guarantees that the total
concentration is conserved in the sense that $d/{dt}\int_V c\left(\bm{x},t\right)d^n
x=0+\text{[boundary terms]}$.  We let $V$ denote the problem domain as well
as the system volume.

The CH equation without flow exhibits dynamical equilibrium in the following
sense.  Starting from a concentration field fluctuating around the unstable
equilibrium $c\left(\bm{x},t\right)=0$, the late-time concentration field has
properties that are time-independent when lengths are measured in units of
typical bubble size $R_{\mathrm{b}}$.  Such properties include the structure
function \cite{Toral_scaling} which we introduce below.

A measure of the correlation of concentration between neighbouring points, given in Fourier space, is the following:
\begin{equation}S\left(\bm{k},t\right) = \frac{1}{V}\int_V d^n x\int_V d^n x' e^{-i\bm{k}\cdot\bm{x}}\left[c\left(\bm{x}+\bm{x'}\right)c\left(\bm{x}\right)-\langle
c\rangle^2\right],\end{equation}
where angle brackets denote the spatial average.  We normalize this
function and compute its spherical average to obtain the structure function
\begin{equation}
s\left(k,t\right)=\frac{1}{\left(2\pi\right)^n}\frac{\tilde{S}\left(k,t\right)}{\langle
c^2\rangle-\langle c\rangle^2}.
\label{eq:structure_fn}
\end{equation}
The spherical average $\tilde{\phi}\left(k\right)$ any function $\phi\left(\bm{k}\right)$
is defined as
\begin{equation}
        \tilde{\phi}\left(k\right) = \frac{1}{\Omega_n}\int d\Omega_n \phi\left(\bm{k}\right),
\end{equation}
where $d\Omega_n$ is the element of solid angle in $n$ dimensions and $\Omega_n$
is its integral.  Thus $\tilde{c}_k$ is the spherical average of the Fourier
coefficient $c_{\bm{k}}$.  For a symmetric binary fluid $\langle c\rangle
= 0$, the structure function is simply the spherically averaged
power spectrum:
\begin{equation}
        s\left(k,t\right)=\frac{1}{\left(2\pi\right)^n}\frac{\left|\tilde{c}_k\right|^2}{\langle
        c^2\rangle},\qquad\text{for }\langle c\rangle=0.
\end{equation}   
The dominant length scale is identified with the reciprocal of the most important
$k$-value, defined as the first moment of the distribution $s\left(k,t\right)$,
\begin{equation}
k_1=\frac{\int_0^\infty k s\left(k,t\right)dk}{\int_0^\infty
s\left(k,t\right)dk}.
\end{equation}
Since the system we study is isotropic except on scales comparable
to the size of the problem domain, this length scale is is a measure of scale
size in each spatial direction.  That is, we have lost no information in
performing the spherical average in Eq.~\eqref{eq:structure_fn}.  This
length scale is in turn identified with the bubble size: $R_{\mathrm{b}}\propto1/k_1$.
  In two-dimensional simulations ($n=2$) it is found \cite{Zhu_numerics,
  Toral_scaling} that while $k_1$ and $s\left(k,t\right)$ depend on time,
  $k_1^n s\left(k/k_1,t\right)$ is a time-independent function with a single
  sharp maximum, confirming that the system is in dynamical
equilibrium and that a dominant scale exists.  Indeed, the growth law $R_{\mathrm{b}}\sim
t^{1/3}$ is obtained, in agreement with the evaporation-condensation picture
of phase separation proposed by Lifshitz and Slyozov (LS exponent)~\cite{LS}.

We present some simple scaling arguments to reproduce the $t^{1/3}$ scaling
law and to
illuminate the effect of flow.  For $\bm{v}=\bm{0}$,
the equilibrium solution of Eq.~\eqref{eq:CH} is $c \approx \pm1$ in domains,
with small transition regions of width $\sqrt{\gamma}$ in between.  Across
these transition regions, it can be shown \cite{LowenTrus} that
\[\mu = -{\Gamma\kappa}/2,\]
where 
\begin{equation}
        \Gamma = \sqrt{{8\gamma}/9}
        \label{eq:integral_st}        
\end{equation}
is the surface tension and $\kappa$ is the radius of curvature.
 Thus, if $R_{\mathrm{b}}$ is a typical bubble size, that is, a length over which $c$ is constant, the chemical potential associated with the bubble is
\begin{equation}\delta\mu\sim\Gamma / R_{\mathrm{b}}.\end{equation}
Balancing terms in Eq.~\eqref{eq:CHmu}, it follows that the time $t$ required
for a bubble to grow to a size $R_{\mathrm{b}}$ is $1/t\sim {\Gamma D}/{R_{\mathrm{b}}^3}$,
implying the LS growth law $R_{\mathrm{b}}\sim \left(\Gamma D t\right)^{1/3}$.

Now the CH free energy $F[c]$ is the system energy owing to the presence of bubbles:
\begin{equation}F[c]=\int_V d^nx \left[\tfrac{1}{4}\left(c^2-1\right)^2
    +\tfrac{1}{2}{\gamma}\left|\nabla
    c\right|^2\right].
\end{equation}
The surface tension $\Gamma$ is the free energy per unit area.  A bubble
carries free energy $\Gamma R_{\mathrm{b}}^{n-1}$, where prefactors due to
angular integration are omitted.  The total free energy $F\left[c\right]$
in the motionless case is then $\Gamma R_{\mathrm{b}}^{n-1} \times N_{\mathrm{b}}$,
where $N_{\mathrm{b}}$ is the total number of bubbles.  Because the system
is isotropic and because there is a well-defined bubble size, we estimate
$N_{\mathrm{b}}$ by $V/{R_{\mathrm{b}}^n}$.  The free energy then has the
scale dependence $F/V\sim\Gamma/R_{\mathrm{b}}$, and using the growth law
for the length $R_{\mathrm{b}}$ we obtain the asymptotic free energy relation
\begin{equation}
  \frac{F}{V}\sim \left(\frac{\Gamma^2}{D}\right)^{1/3}t^{-1/3}.
  \label{eq:fe_decay}
\end{equation}
By introducing the tracer variance
\begin{equation}
\sigma^2=\langle c^2\rangle-\langle c\rangle^2
\label{eq:sigsq}
\end{equation}
and restricting to a symmetric mixture $\langle c \rangle = 0$, we may identify
$\sigma^2\simeq V_{\mathrm{b}}/V$, where $V_{\mathrm{b}} = N_{\mathrm{b}}
R_{\mathrm{b}}^n$ is the volume occupied by bubbles.  Since $F=\Gamma R_{\mathrm{b}}^{n-1}
N_{\mathrm{b}}$, it follows that
\begin{equation}
        R_{\mathrm{b}}\sim\sigma^2 / F.
        \label{eq:Rdsig}
\end{equation}
We shall verify these results in Section~\ref{sec:numerics}.
Equation~\eqref{eq:Rdsig} will provide a useful measure of bubble size when a
flow is imposed, since no assumption is made about the number of bubbles per
unit volume.  In a dynamical scaling regime, the length scales calculated from
these energy considerations must agree with that computed from the first
moment of the structure function, since there is only one length scale in such
a regime \cite{Furukawa_scales}.

Stirring a CH fluid introduces new length scales.  The flow will alter the
sharp power spectrum found above and so it may not be possible to extract
definite scales from experiments or numerical simulations.  We might expect
further ambiguity of scales during a regime change (crossover), for example, as the bubbles are broken up and diffusion takes over.  Nevertheless, for a given flow it is possible to construct length scales from
the system parameters.  We shall impose a flow $\bm{v}\left(\bm{x},t\right)$
that is chaotic in the sense that nearby particle trajectories separate
exponentially in time at a mean rate given by the Lyapunov exponent.

If the advection term and the surface tension term have the same order of
magnitude in a chaotic flow, by balancing the terms $\bm{v}\cdot\nabla c$
and $D\nabla^2\mu$ we obtain a scale
\begin{equation}
  R_{\mathrm{st}} = \left(\Gamma D / \lambda\right)^{1/3},
  \label{eq:Rdsim}
\end{equation}
where $\lambda$ is the Lyapunov exponent of the flow.  Because the term
$\bm{v}\cdot\nabla c$ gives an exponential amplification of gradients, the
balance of segregation and hyperdiffusion in the term $\mu = f_0'\left(c\right)-\gamma\nabla^2c$
may be broken and hyperdiffusion may overcome segregration.  This will happen
on a scale~$R_{\mathrm{diff}}$ given by the balance of the terms $\bm{v}\cdot\nabla
c$ and $\gamma\nabla^4c$,
\begin{equation}
        R_{\mathrm{diff}}=\left({\gamma D} / \lambda \right)^{1/4}.
        \label{eq:R_diff}        
\end{equation}
In this regime, we expect mixing owing to the presence of the advection and the (hyper) diffusion.  Using Eqs.~\eqref{eq:integral_st},~\eqref{eq:Rdsim}
and~\eqref{eq:R_diff},
the crossover between the bubbly and the diffusive regimes takes place when
$\lambda\simeq D/\gamma$. 
In the following sections we shall examine these scales and look for this
crossover between bubbles and filaments, the latter being characteristic
of mixing by advection-diffusion.
\section{Numerical Solution of the Two-Dimensional Problem}
\label{sec:numerics}
In order to run simulations at high resolution, we specialize to two
dimensions.  We solve Eq.~\eqref{eq:CH} with and without flow to investigate
the scaling laws presented in the previous section.

Equation~\eqref{eq:CH} is integrated by an operator splitting technique, in
which a specialized advective step followed by a phase-separation step are
performed at each iteration.  The phase separation step is the semi-implicit
spectral algorithm proposed by Zhu et al.~\cite{Zhu_numerics} for the case
without flow.  The semi-implicitness of the algorithm enables us to use a
reasonably large time step.  We use the following velocity field,
defined on a $\left(2\pi\right)\times\left(2\pi\right)$ grid of $512^2$ points
with periodic boundary conditions,
\begin{equation}
\begin{split}
  v_x\left(x,y,t\right) &= \alpha\sin\left(y+\phi_n\right),\qquad v_y=0,\qquad
  n\tau\leq t<\left(n+\tfrac{1}{2}\right)\tau,\\
  v_y\left(x,y,t\right) &= \alpha\sin\left(x+\psi_n\right),\qquad v_x=0,\qquad
  \left(n+\tfrac{1}{2}\right)\tau \leq t<\left(n+1\right)\tau,
\end{split}
\label{eq:sineflow}
\end{equation}
where $\phi_n$ and $\psi_n$ are phases that are randomized once during each
flow period $\tau$ and the integer $n$ labels the period.  We set
$\tau=1$ and work in units where time is the number of periods.  We use Pierrehumbert's
lattice method for advection~\cite{lattice_PH1, lattice_PH2}.  With the problem
defined on a discrete grid, this method is exact for the advection step. 
Using standard techniques~\cite{Eckmann1985,WigginsIntroDynSys}, we compute
the Lyapunov exponent numerically and find it is always positive and independent
of initial conditions, as expected for this flow.  For the parameter regime
$\alpha\in\left[0,1\right]$ we find
\begin{equation}
  \lambda \sim 0.118\phantom{l}\alpha^2,\qquad \alpha \lesssim 1,
\end{equation}
a result that we shall use in what follows.  For sufficiently large
$\alpha$, it is easy to show the exact asymptotic result
\begin{equation}
  \lambda \sim \log\left(\tfrac{1}{4}\,{\alpha^2}\right),
  \qquad\alpha\gg1.
\end{equation}
Furthermore, the correlation length for the velocity is of order of the box
size, which is the largest scale in the problem.  Therefore, using the
prediction of Lacasta et al.~\cite{Sancho}, we expect this velocity field
to produce coarsening arrest even at very small stirring amplitudes.

We integrate Eq.~\eqref{eq:CH} in the manner explained above.  The initial
conditions are chosen to represent the sudden cooling of the binary
fluid: Above a certain temperature $T_{\mathrm{c}}$ the homogeneous state
$c=0$ is stable, while below this temperature the mixture free energy changes
character to become $\tfrac{1}{4}\left(c^2-1\right)^2$, which makes the homogeneous
state unstable.  Thus, at temperatures $T>T_{\mathrm{c}}$, $c\left(\bm{x}\right)=0+
\text{[fluctuations]}$, and the sudden cooling of the system below $T_{\mathrm{c}}$
%
%
\begin{figure}[htb]
\subfigure[]{
  \scalebox{0.50}[0.50]{\includegraphics*[viewport=0 0 440 330]{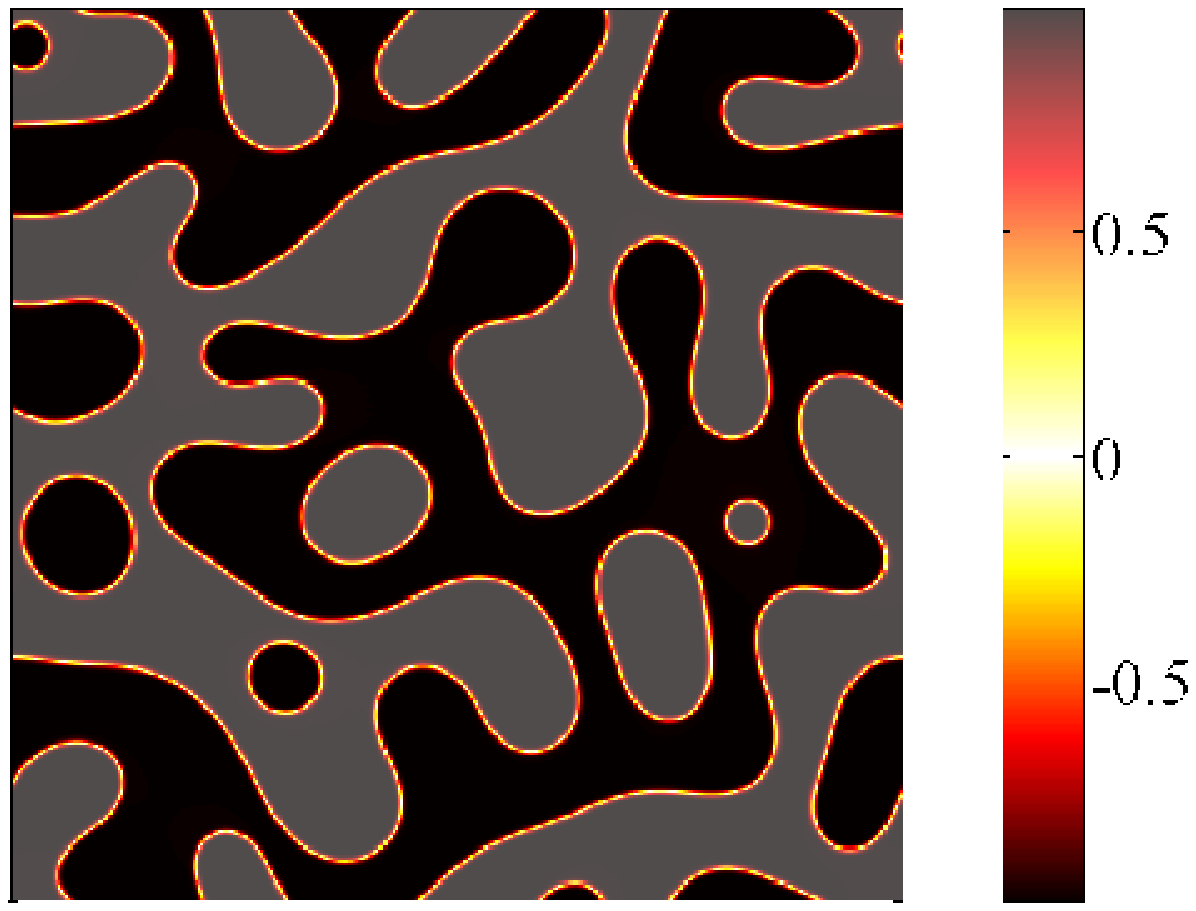}}
}
\subfigure[]{
   \scalebox{0.50}[0.50]{\includegraphics*[viewport=0 0 420 330]{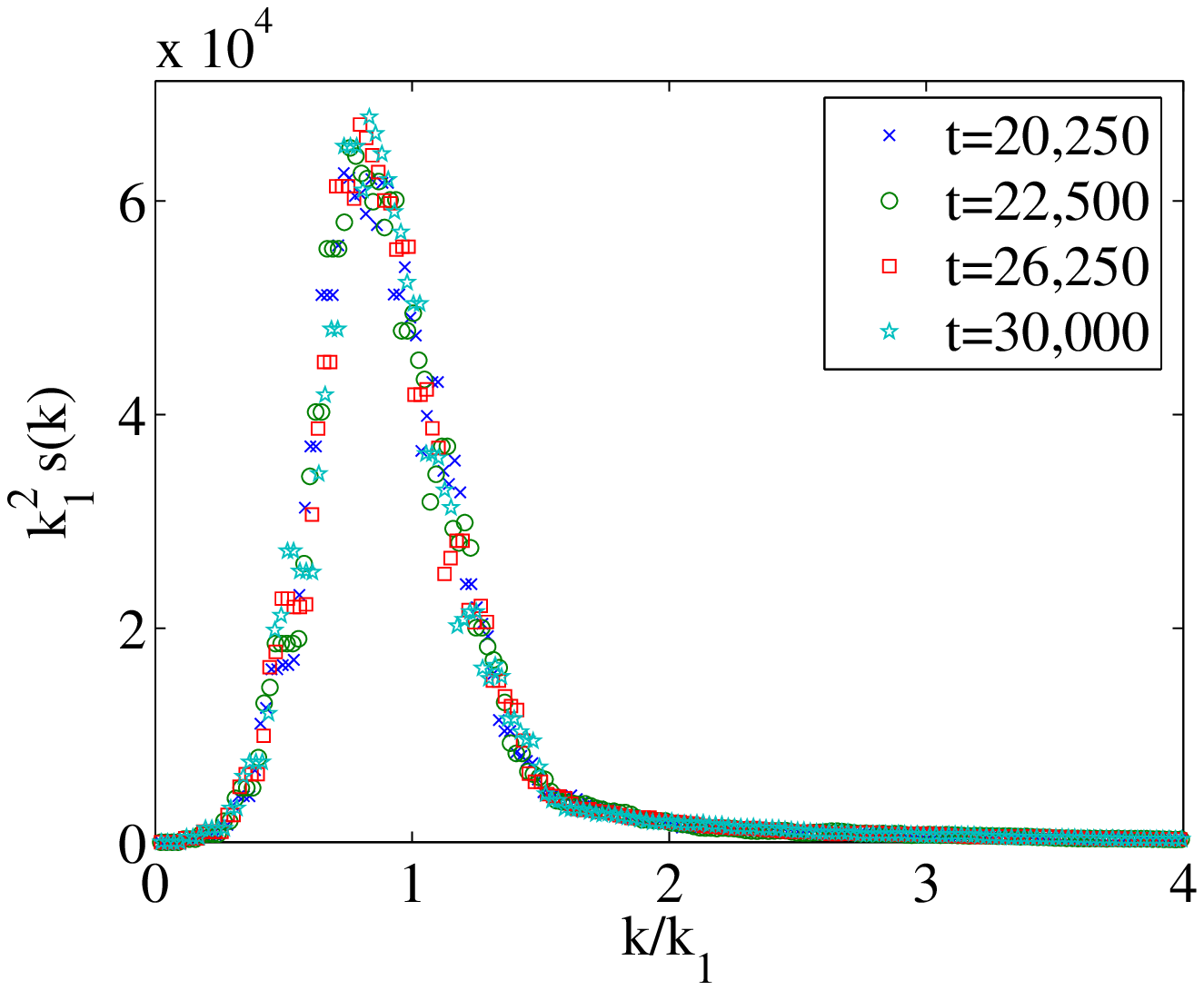}}
}
\subfigure[]{
   \scalebox{0.50}[0.50]{\includegraphics*[viewport=0 0 420 330]{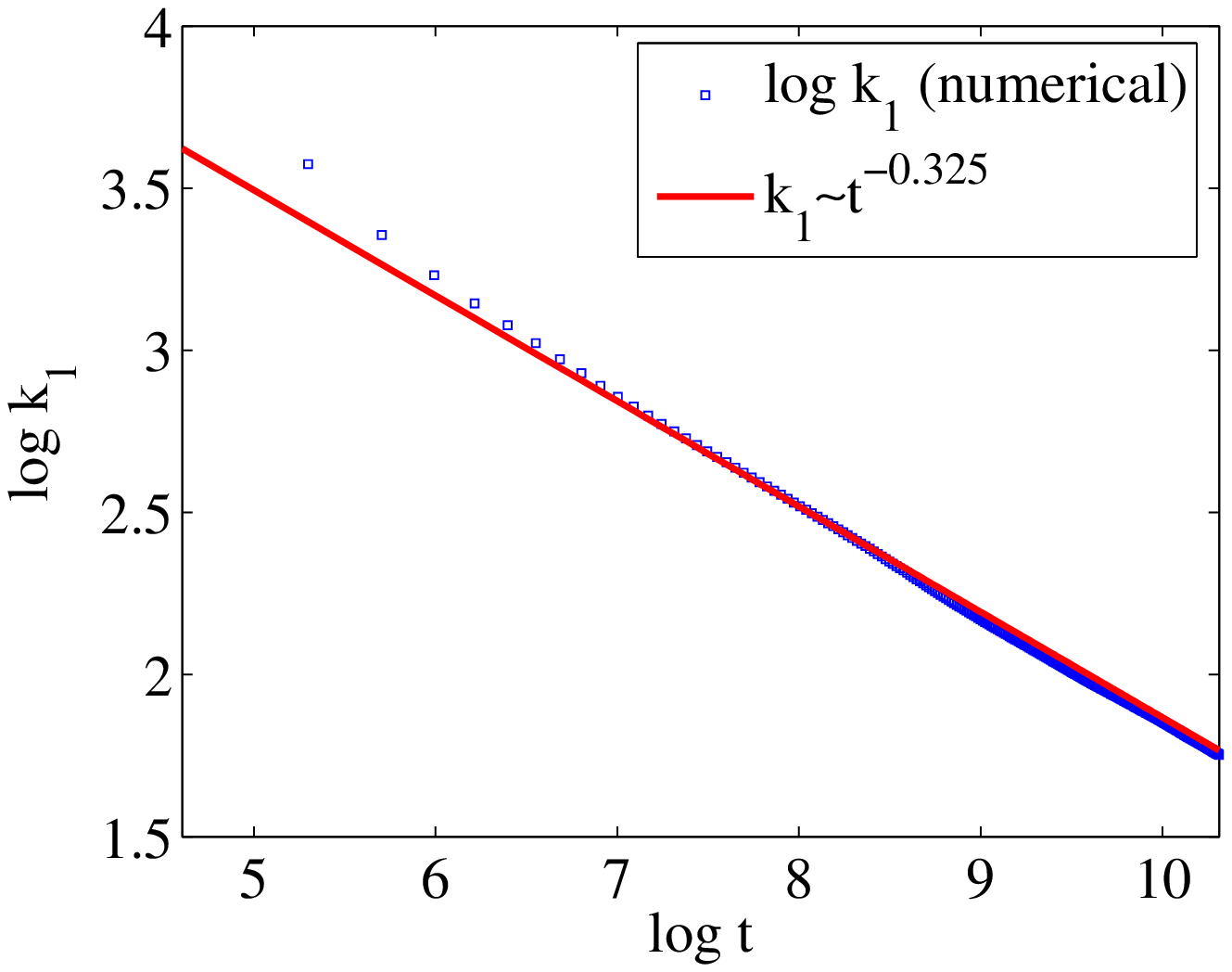}}
}
\subfigure[]{
   \scalebox{0.50}[0.50]{\includegraphics*[viewport=0 0 420 330]{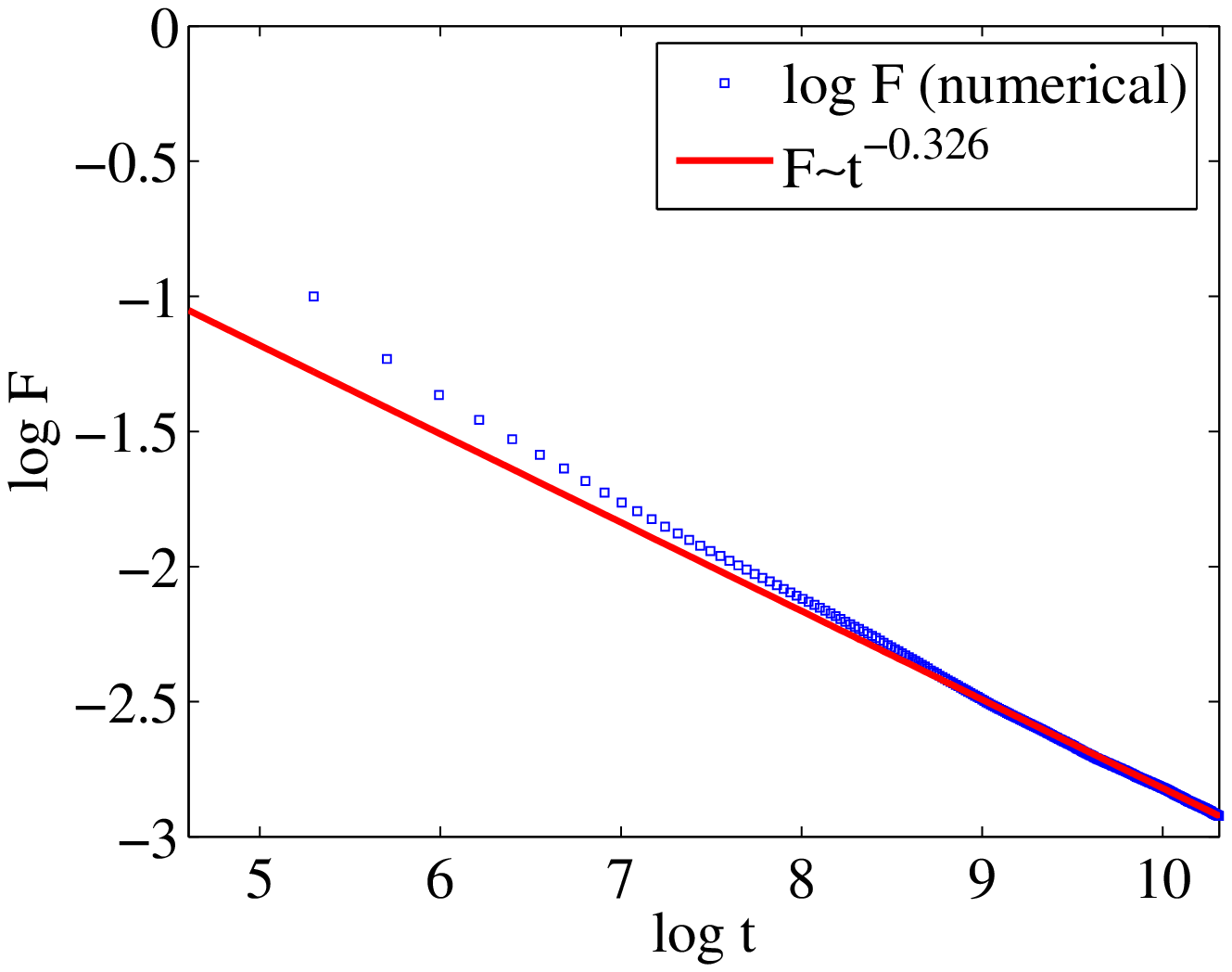}}
}
\caption{%
(a) Concentration, at $t=30,000$;
(b) Structure function $k_1^2 s\left(k/k_1,t\right)$ in the scaling regime
$t\geq20,000$; (c) Time dependence of $k_1$, with power law exponent close
to the LS exponent $1/3$;
(d) Free energy, $F$, exhibiting the time dependence $F\sim t^{-0.326}$,
close to the LS exponent.
}
\label{fig:CHnoflow}
\end{figure}
leaves the system in this state.  In order for the CH equation without thermal
noise~\eqref{eq:CH} to describe the evolution of these initial conditions,
the cooling we have imposed must take the system to $T=0$.  Thus, we are
working in the so-called deep-quench limit.  With the initial conditions
$c\left(\bm{x},0\right)=0+\text{[fluctuations]}$, both components of the
fluid are present in equal amounts and the spatial average of the concentration
field is zero.
 
Although the case without flow has been investigated before \cite{Zhu_numerics,
Toral_scaling}, we reproduce it for two reasons.  First, it serves as a validation
of our algorithm.  Second, we shall confirm the free energy
laws~\eqref{eq:fe_decay} and~\eqref{eq:Rdsig} which to our knowledge have
not been examined before. Finding an appropriate measure of bubble growth
will be important for understanding the behavior of the bubbles when we introduce
stirring.

The results of these simulations are presented in Fig.~\ref{fig:CHnoflow}.
 The scaling function $k_1^2s\left(k/k_1,t\right)$ is approximately time-independent
 for $t>20,000$ as evidenced by Fig.~\ref{fig:CHnoflow}(b), implying
 that the scaling exponents for $k_1$ and $F$ are to be extracted from the
 late-time data with $t>20,000$.  For $t>30,000$ finite-size effects spoil
 the scaling laws.  Thus while the scaling exponents are extracted from a
 small window, the fit is good and this suggests that the power laws obtained
 give the true time-dependence of $k_1$ and $F$ at late times.  In this way
 we recover the dynamical scaling regime in which $k_1^2 s\left(k/k_1,t\right)$
 is time-independent, in addition to the decay law $k_1\sim t^{-1/3}$ \cite{Zhu_numerics}.
  Running the simulation repeatedly for an ensemble of random initial conditions
  improves the accuracy of the exponent.  Furthermore we obtain the energy
  laws $F\sim c_1 t^{-1/3}$, and $1-\sigma^2\sim c_2t^{-1/3}$.  Thus,
\begin{equation}
\frac{\sigma^2}{F}=\frac{1-c_2 t^{-1/3}}{c_1t^{-1/3}}\sim\frac{1}{c_1}t^{1/3},\qquad\text{for
}t\gg1
\end{equation}
so that $\sigma^2/F$, $1/F$ and $1/k_1$ all grow as $t^{1/3}$ and hence provide
identical measures of scale growth, in agreement with the classical assumption
that there is a unique length scale in the problem \cite{Bray_advphys, Zhu_numerics}.

\begin{figure}[htb]
\centering
\noindent
%
\subfigure[]{\scalebox{0.50}[0.50]{\includegraphics*[viewport=0 0 400 330]{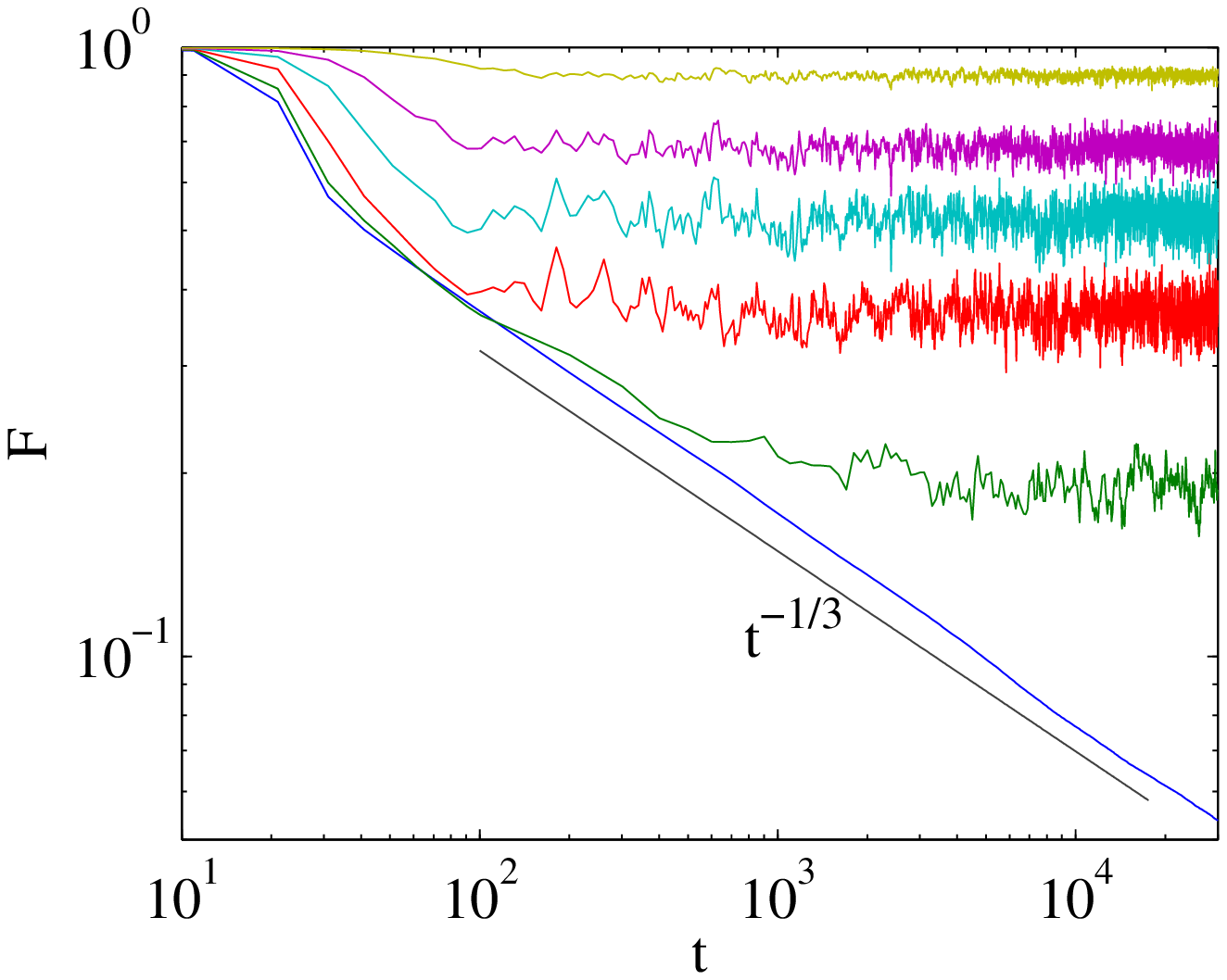}}}
%
\subfigure[]{\scalebox{0.50}[0.50]{\includegraphics*[viewport=0 0 400 330]{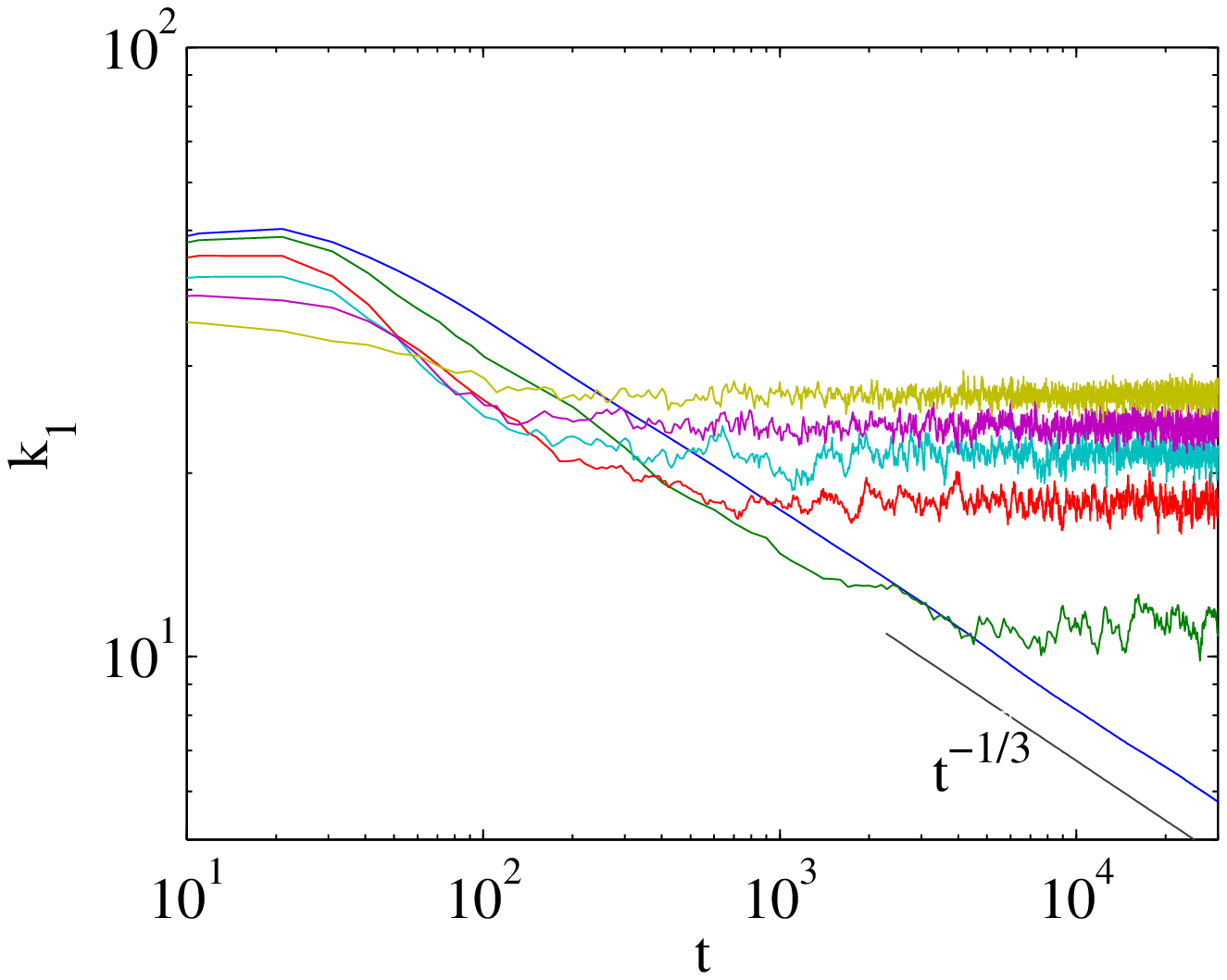}}}
%
\subfigure[]{\scalebox{0.50}[0.50]{\includegraphics*[viewport=0 0 400 330]{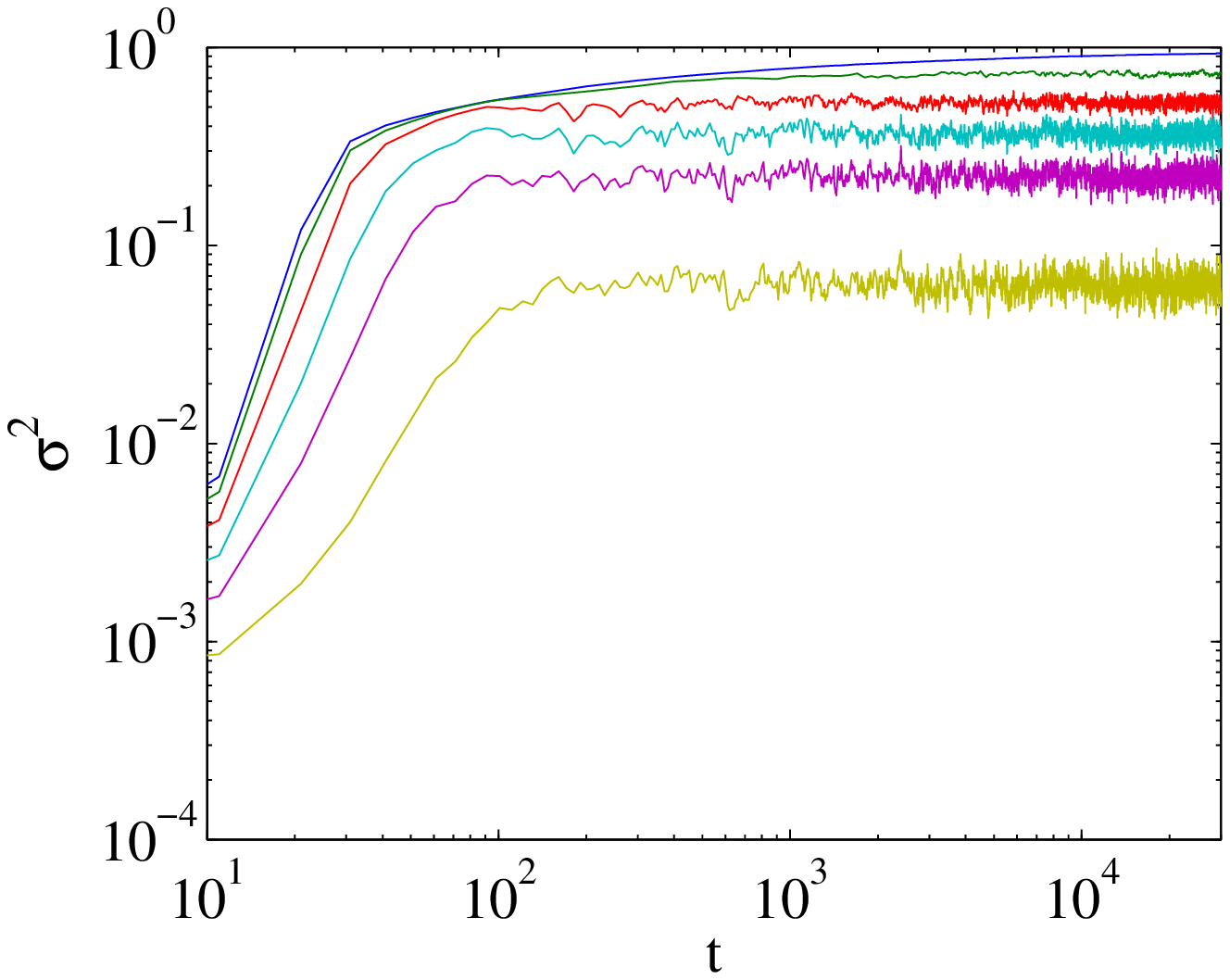}}}
\caption{(a) $F$ vs $t$ for (from bottom) $\alpha=0,0.1,0.3,0.5,0.7,1.0$;
(b) $k_1$ vs $t$ for the same values of $\alpha$; (c) Tracer variance $\sigma^2$
for (from bottom) $\alpha = 1.0,0.7,0.5,0.3,0.1,0$.}
\label{fig:fluctuations}
\end{figure}

We investigate the stirred case in a similar manner and obtain the results
below, after $t=30,000$ time steps.  In order to ensure that we are in a
steady state, we study the inverse lengths $F$ and $k_1$
and the tracer variance $\sigma^2$ (see Eq.~\eqref{eq:sigsq}) which measures the homogeneity of the concentration ($\sigma^2=0$ for a homogeneous
mixture)~\cite{lattice_PH1,lattice_PH2}.  These quantities are time-dependent and have the property that after some transience, they fluctuate around a mean value, without secular trend.  This can be seen in Fig.~\ref{fig:fluctuations}.

\begin{figure}[htb]
\centering
\noindent
\subfigure[]{\scalebox{0.45}[0.45]{\includegraphics*[viewport=50 0 420 330]{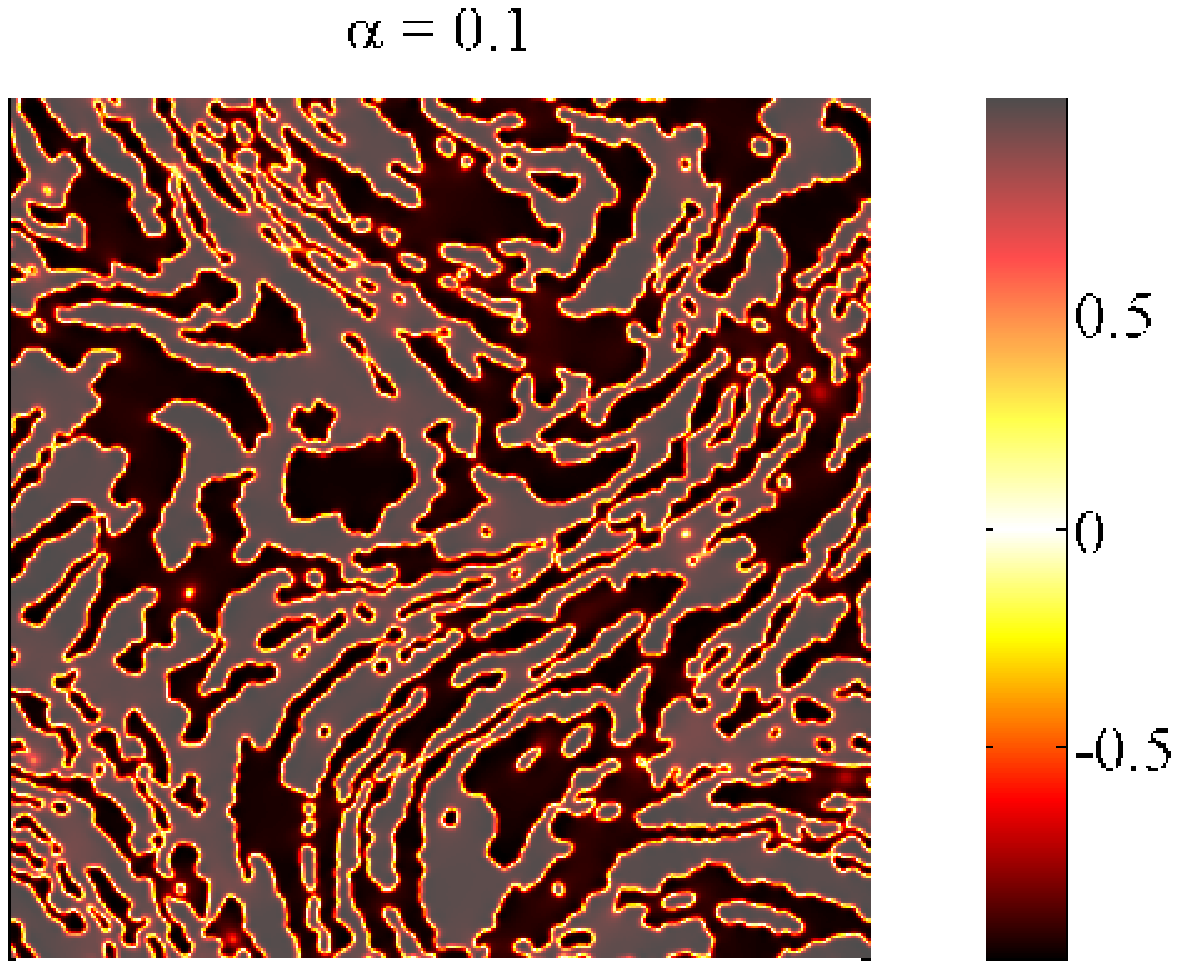}}}
%
\subfigure[]{\scalebox{0.45}[0.45]{\includegraphics*[viewport=50 0 420 330]{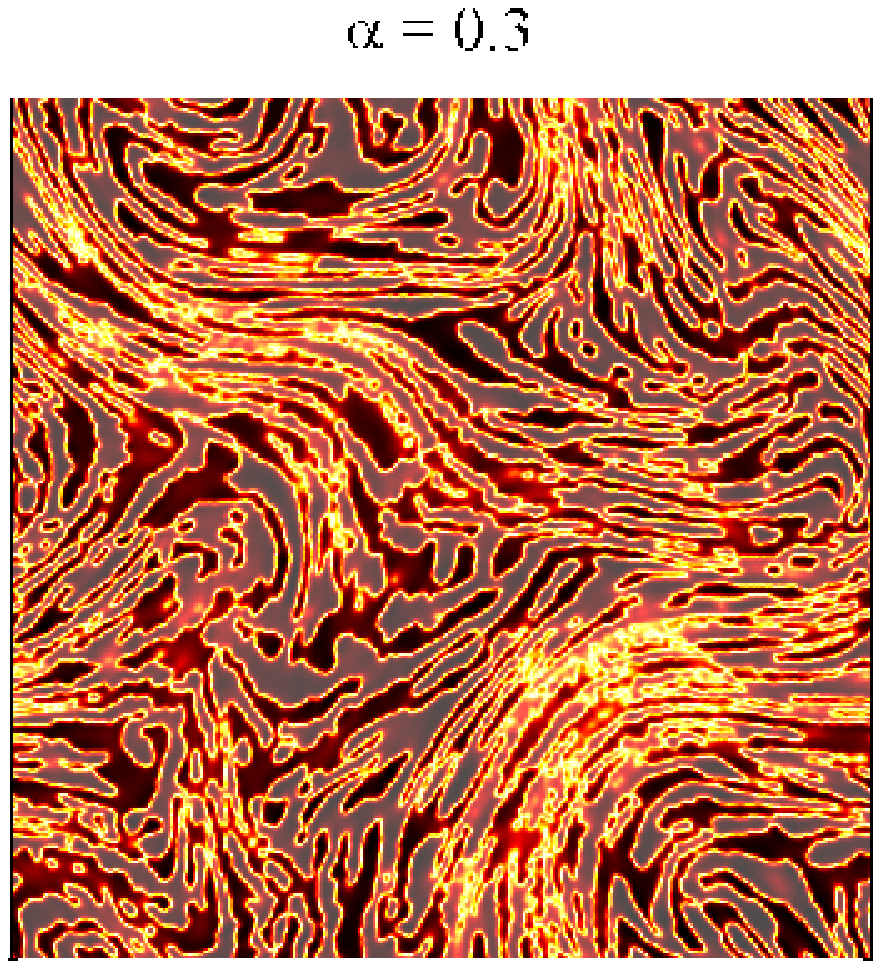}}}
%
\subfigure[]{\scalebox{0.45}[0.45]{\includegraphics*[viewport=50 0 420 330]{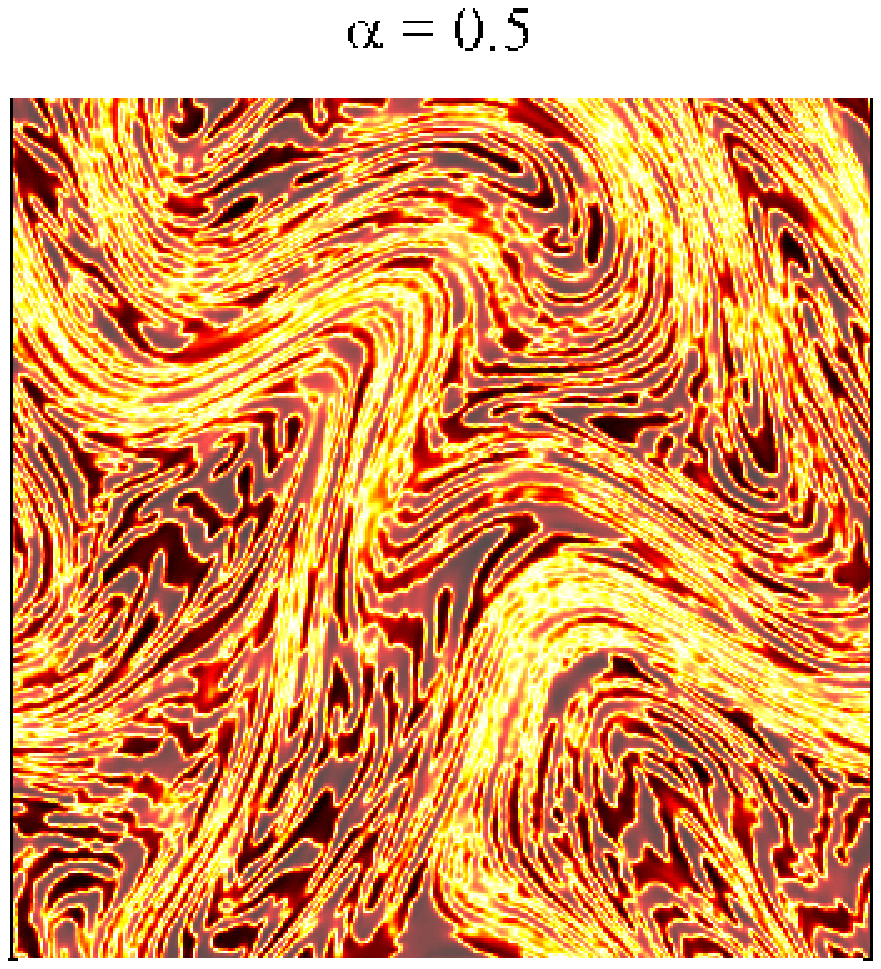}}}
%
\subfigure[]{\scalebox{0.45}[0.45]{\includegraphics*[viewport=50 0 420 330]{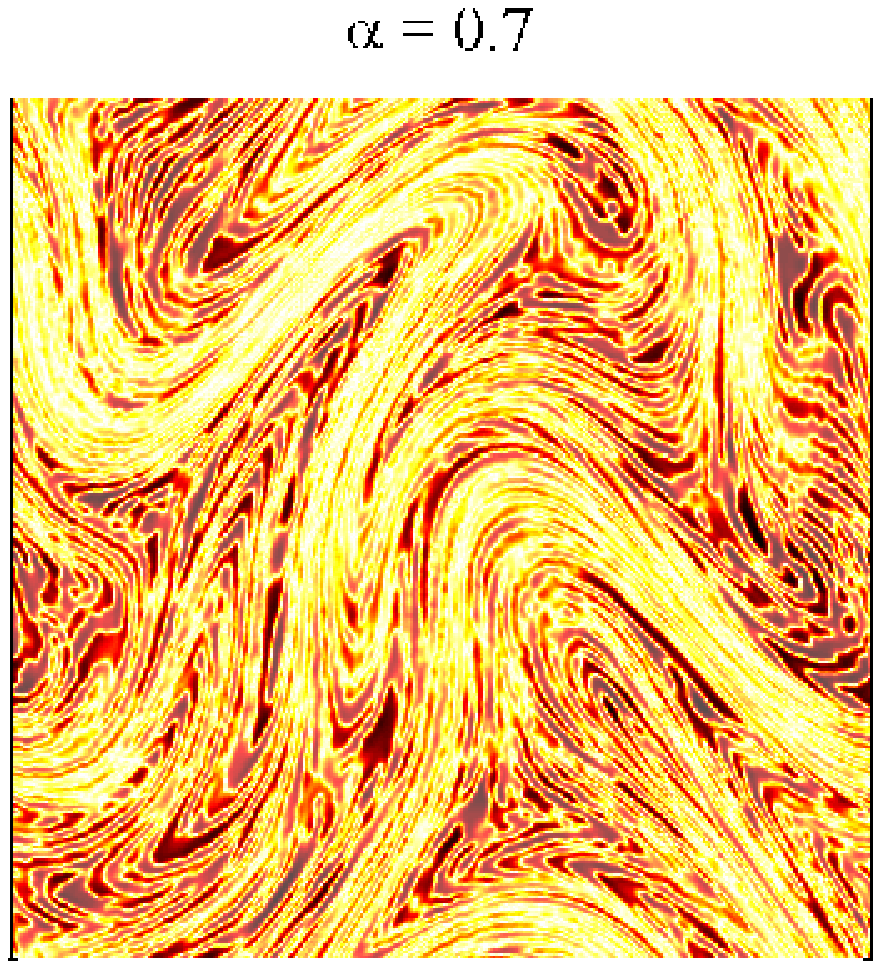}}}
%
\subfigure[]{\scalebox{0.45}[0.45]{\includegraphics*[viewport=50 0 420 330]{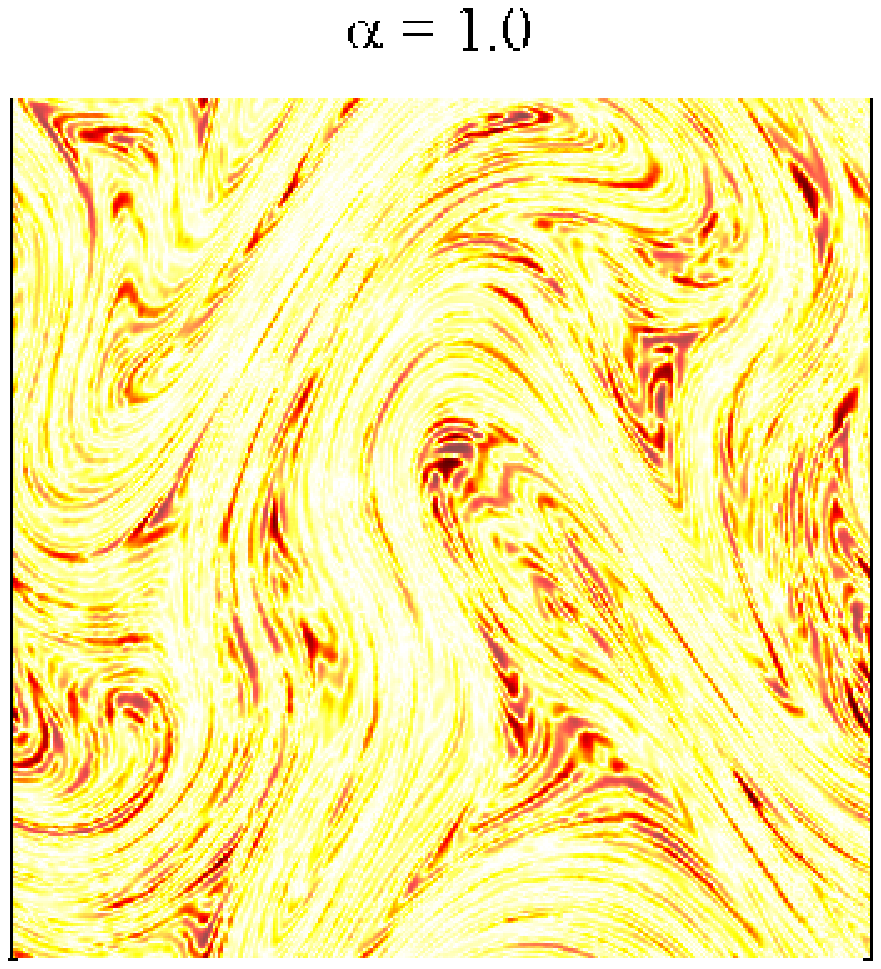}}}
\caption{The steady-state concentration field after $t=30,000$ timesteps,
for various stirring amplitudes $\alpha$.}
\label{fig:CHflow}
\end{figure}

In Fig.~\ref{fig:CHflow}, for $\alpha = 0.1$, the concentration field at late
time looks
%
%
similar to that for $\alpha = 0$.  This is because for small $\alpha$,
the effect of advection is  diffusive, with small movements of particles
giving rise to bubbles with jagged boundaries, but no breakup.  The PDFs
of the concentration for $\alpha\in [0,0.01,...,0.2]$ have the same structure.
 Nevertheless, it is clear from Fig.~\ref{fig:fluctuations}(b) that coarsening
 arrest takes place for $\alpha = [0.01,...,0.2]$, in contrast to $\alpha=0$.
  As $\alpha$ increases, the bubbles become less and less evident.  For large
  $\alpha$, we see regions containing filaments of similar concentration.
   After a period of transience, the free energy $F$, the mean wavenumber
   $k_1$ and the variance $\sigma^2$ fluctuate around mean values, confirming
   the existence of the steady state.  The bubble size is extracted from
   the quantity $\overline{\sigma^2/F}$, where the overbar denotes a time
   average over the fluctuations.  We investigate how the mean lengths scale
   with the Lyapunov exponent $\lambda$.

\begin{figure}[htb]
\centering
\subfigure[]{
  \scalebox{0.5}[0.5]{\includegraphics*[viewport=0 0 400 330]{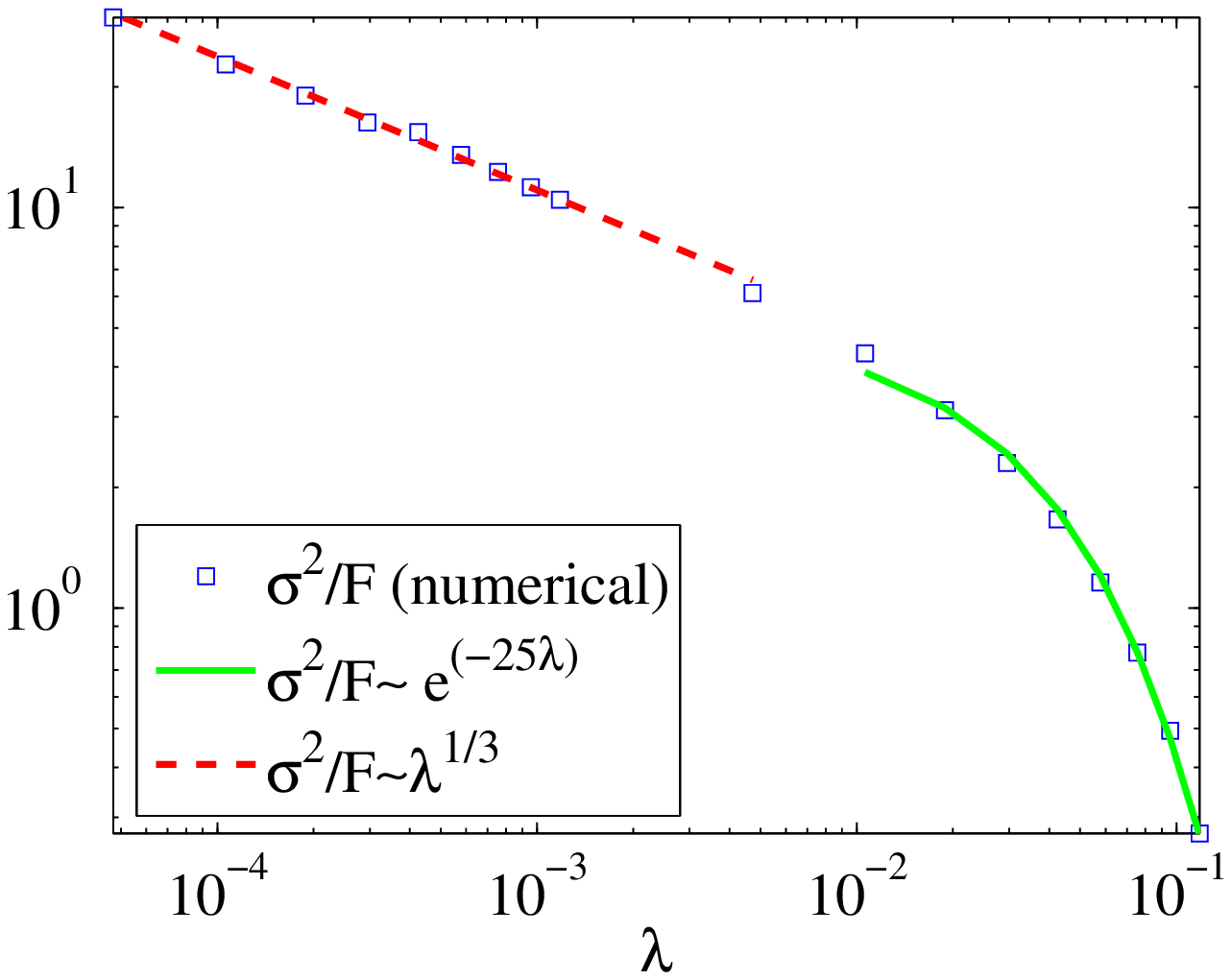}}
}
\subfigure[]{
  \scalebox{0.5}[0.5]{\includegraphics*[viewport=0 0 400 330]{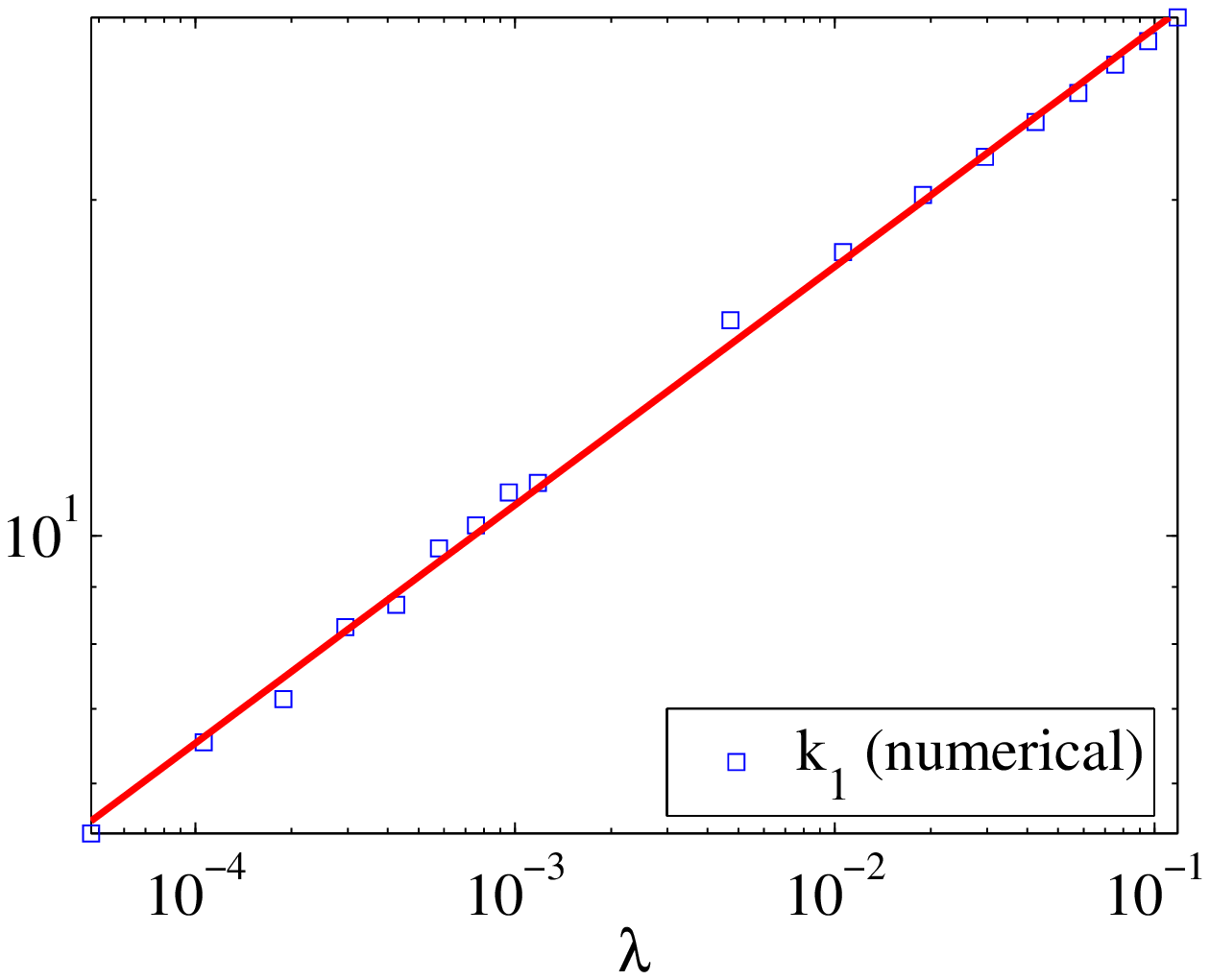}}
}
\subfigure[]{
  \scalebox{0.5}[0.5]{\includegraphics*[viewport=0 0 400 330]{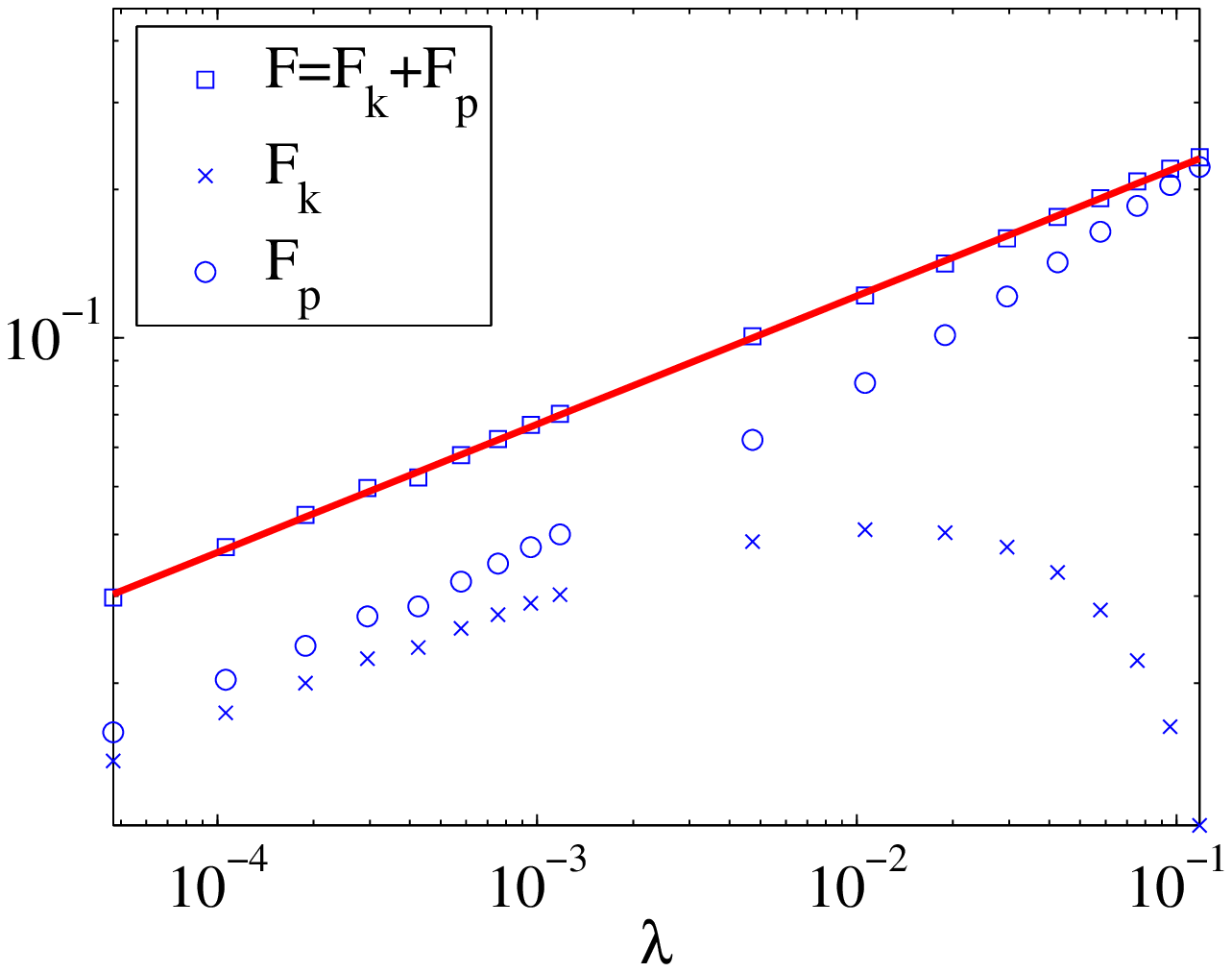}}
}
\subfigure[]{
  \scalebox{0.5}[0.5]{\includegraphics*[viewport=0 0 400 330]{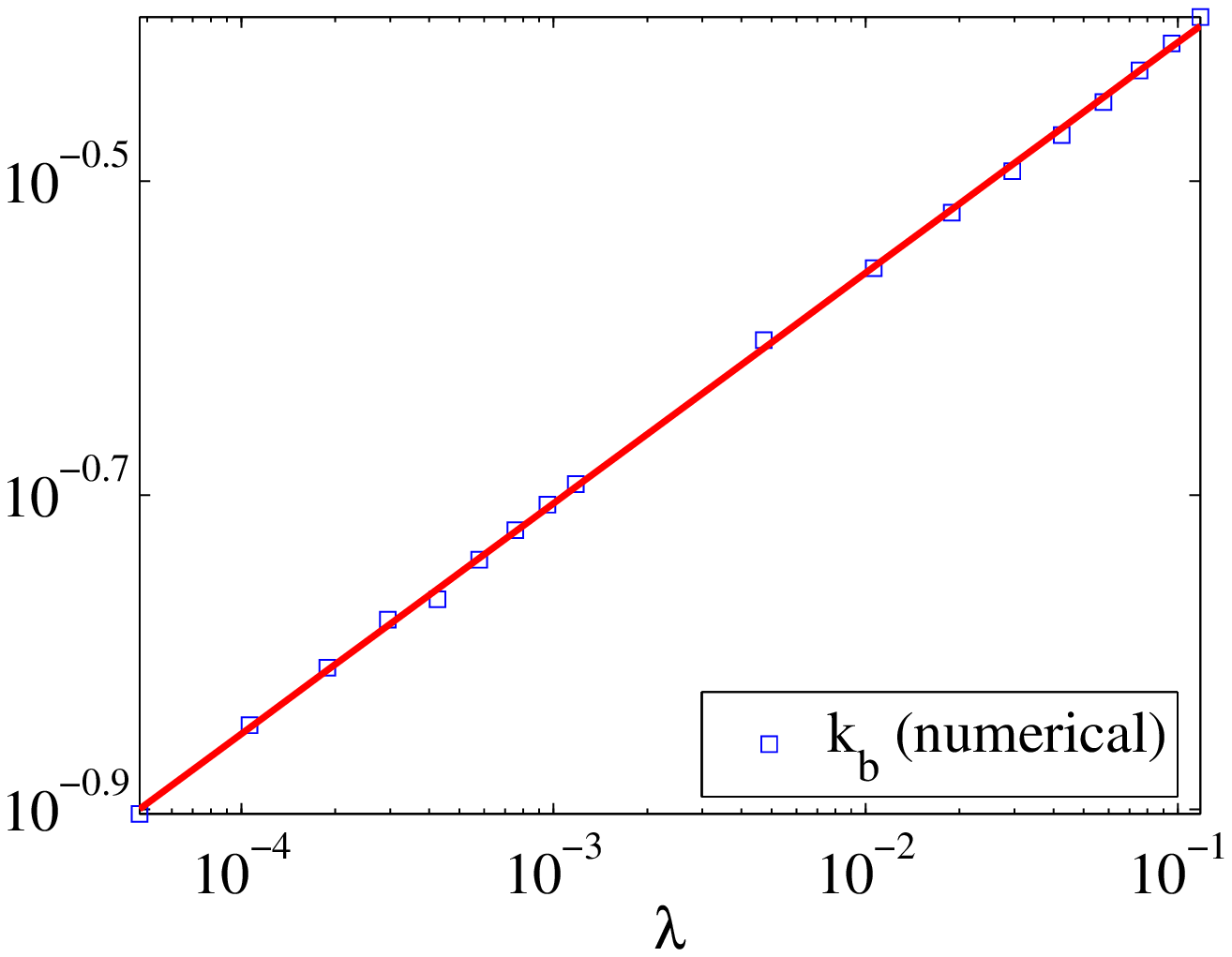}}
}
\caption{(a) 
The plot of $\log\big(\overline{\sigma^2/F}\big)$ against $\log\lambda$ suggests
the scaling law $\overline{\sigma^2/F}\sim\lambda^{-1/3}$ for small stirring
amplitudes while for large amplitudes this proxy for the bubble radius decays
exponentially with increasing Lyapunov exponent.
(b) Graph of $\log\overline{k_1}$ against $\log\lambda$, with approximate
power law behavior $\overline{k_1}\sim \lambda^{0.21}$.  The exponent is
not the same over the entire range, however.
(c) The graph of $\log\overline{F}$ exhibits a much cleaner power law behaviour
$\overline{F}\sim\lambda^{0.26}$.
(d)  The Batchelor wavenumber $k_b=\big[{\langle\left|\nabla c\right|^2\rangle}/{\langle
c^2\rangle}\big]^{1/2}$ also possesses a clear power law behaviour $\overline{ k_b}\sim\lambda^{0.15}$.
}
\label{fig:CHscale}
\end{figure}

In Fig.~\ref{fig:CHscale} we see that for small $\lambda$ the proxy $\overline{\sigma^2/F}$
for bubble radius scales with the
%
%
LS exponent as $\lambda^{-1/3}$, suggesting an equilibrium bubble size on
this scale.  For larger $\lambda$, the hyperdiffusion breaks up these bubbles
and mixes the fluid.  This process leads to a faster than algebraic decay
of bubble size, visible in the figure.  Other measures of the bubble size,
such as the mean wavenumber $\overline{k_1}$ or the quantity $1-\overline{\sigma^2}$
do not possess a clear scaling law.  For example, in Fig.~\ref{fig:CHscale}(b)
we plot $\log\overline{k_1}$ against $\log\lambda$.  The exponent changes
over the three decades of data, indicating that the wavenumber $\overline{k_1}$
does not have a clear scaling law with the Lyapunov exponent $\lambda$. 
However, the dependence of the time-averaged
free energy on the Lyapunov exponent has the power law behavior $\overline{F}\sim\lambda^{0.26}$
over three decades, suggesting a genuine power law relationship.  The Batchelor
wavenumber  $k_b=[{\langle\left|\nabla c\right|^2\rangle}/{\langle
c^2\rangle}]^{1/2}$~\cite{Batchelor1959} also possesses a clean power law
over three decades.

Furukawa~\cite{Furukawa_scales} shows that in a situation of dynamical equilibrium
characterized by a single length scale, the potential part of the free energy
$F_{\mathrm{p}}=\tfrac{1}{4}\int d^2 x\left(c^2-1\right)^2$ should be in
equipartition with the kinetic part $F_{\mathrm{k}}=\tfrac{1}{2}\int d^2x\gamma\left|\nabla
c\right|^2$.  In Fig.~\ref{fig:CHscale} we see that this is not the case
when the bubbles start to break up as the stirring is increased.  This is
an indication of a crossover between the bubbly regime and the well-mixed
one in which the variance is reduced by stirring.

\begin{figure}[htb]
\subfigure[]
{
     \scalebox{0.5}[0.5]{\includegraphics*[viewport=0 0 450 330]{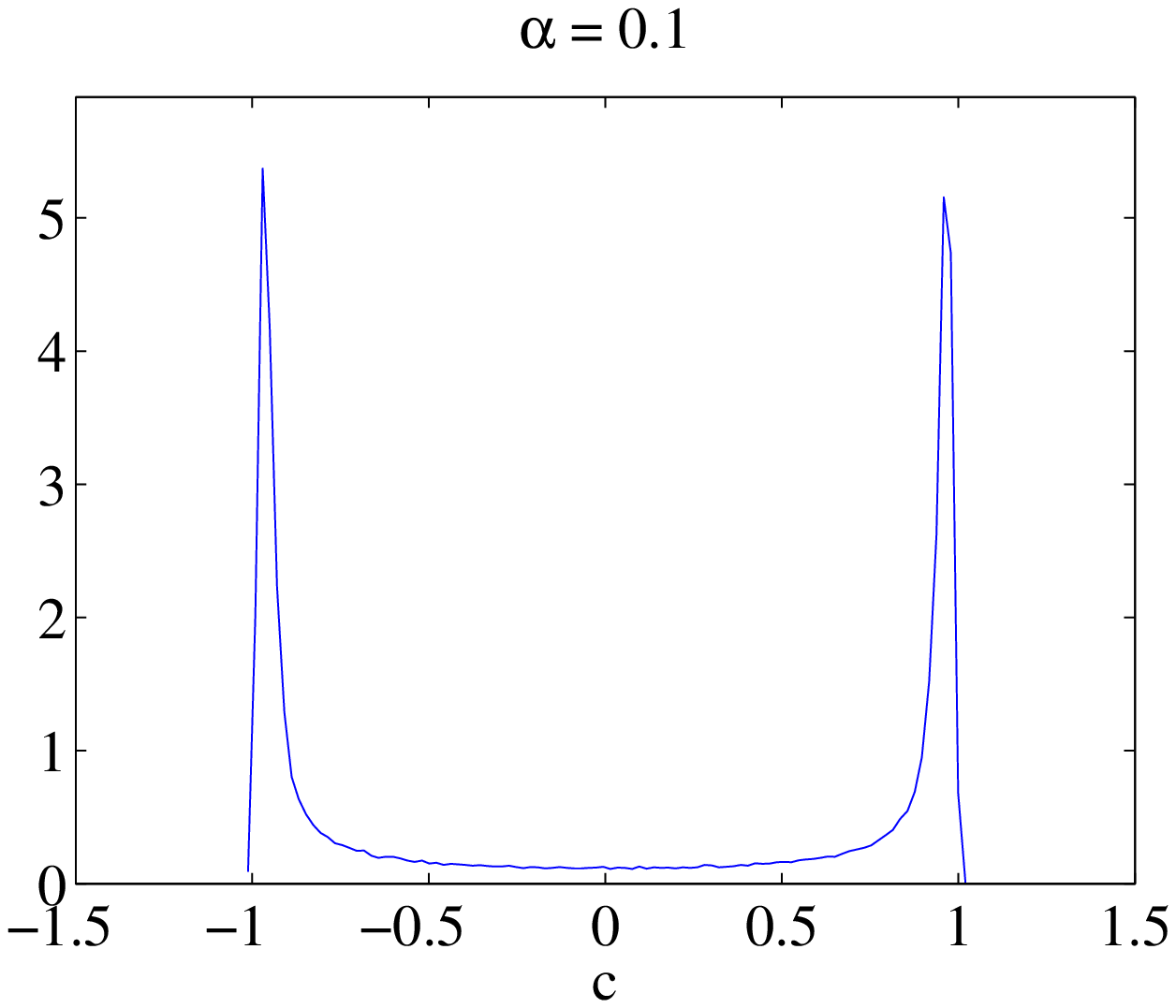}}}
\subfigure[]
{
     \scalebox{0.5}[0.5]{\includegraphics*[viewport=0 0 450 330]{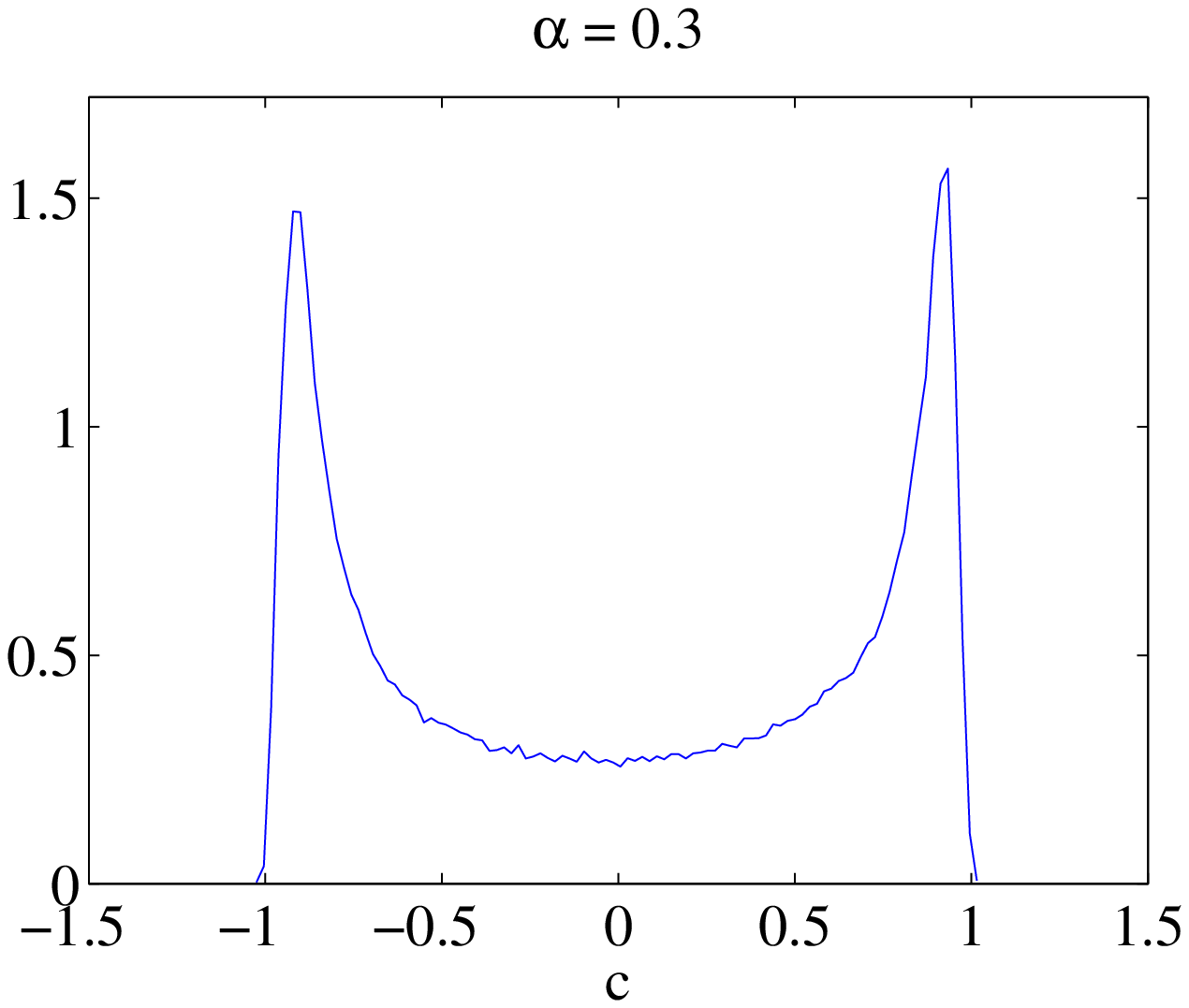}}}
\subfigure[]
{
     \scalebox{0.5}[0.5]{\includegraphics*[viewport=0 0 450 330]{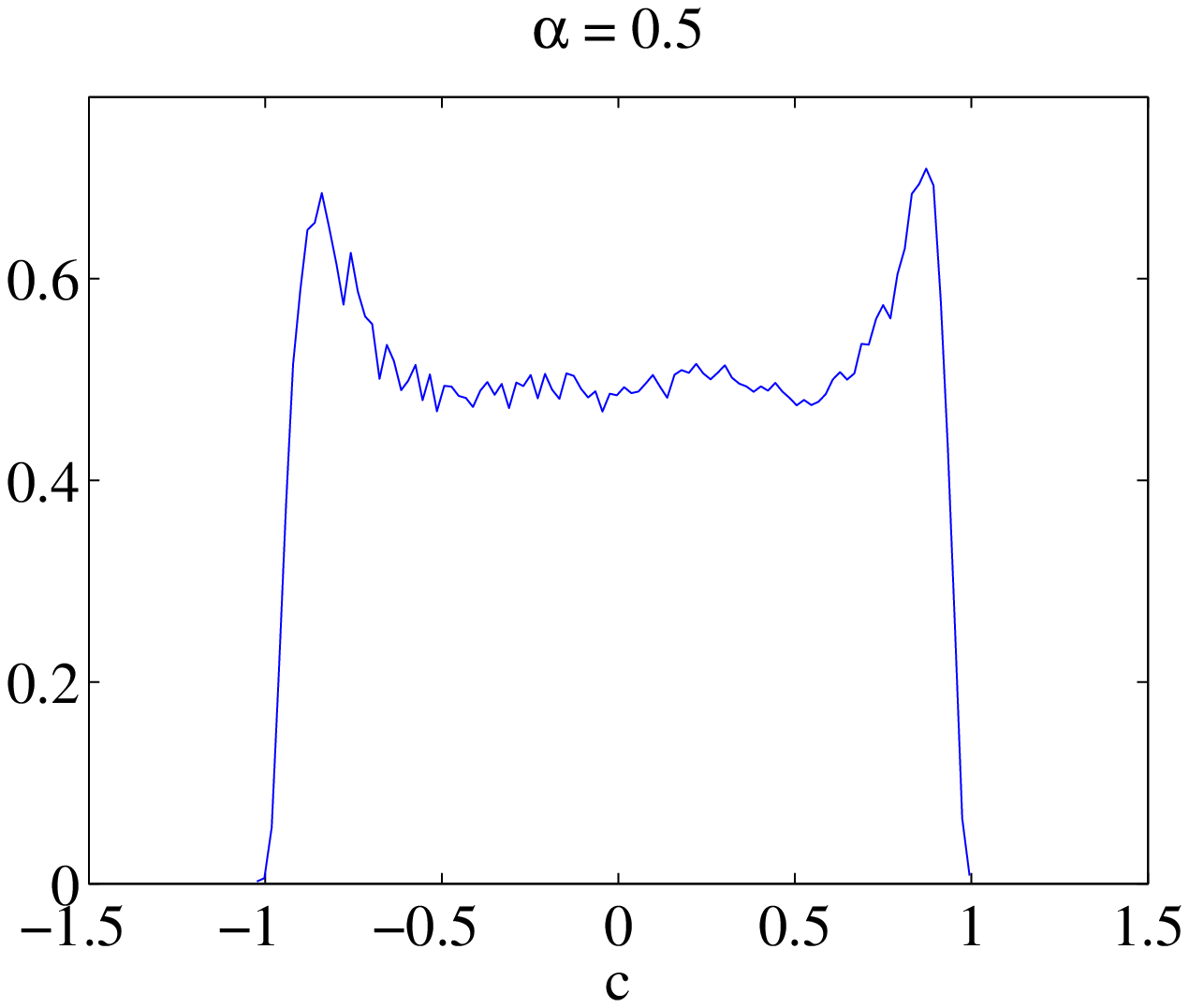}}}
\subfigure[]
{
     \scalebox{0.5}[0.5]{\includegraphics*[viewport=0 0 450 330]{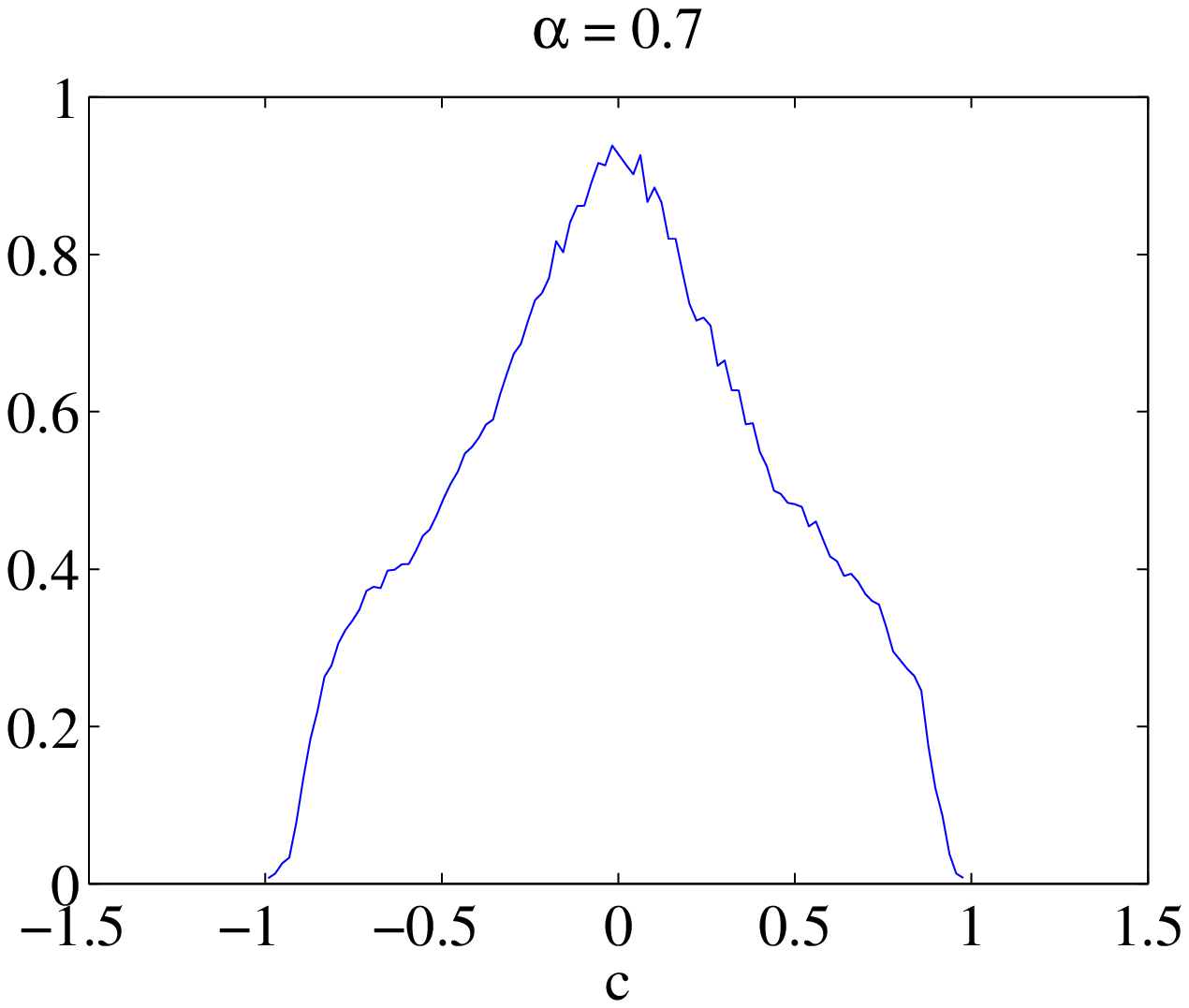}}}
\subfigure[]
{
     \scalebox{0.5}[0.5]{\includegraphics*[viewport=0 0 450 330]{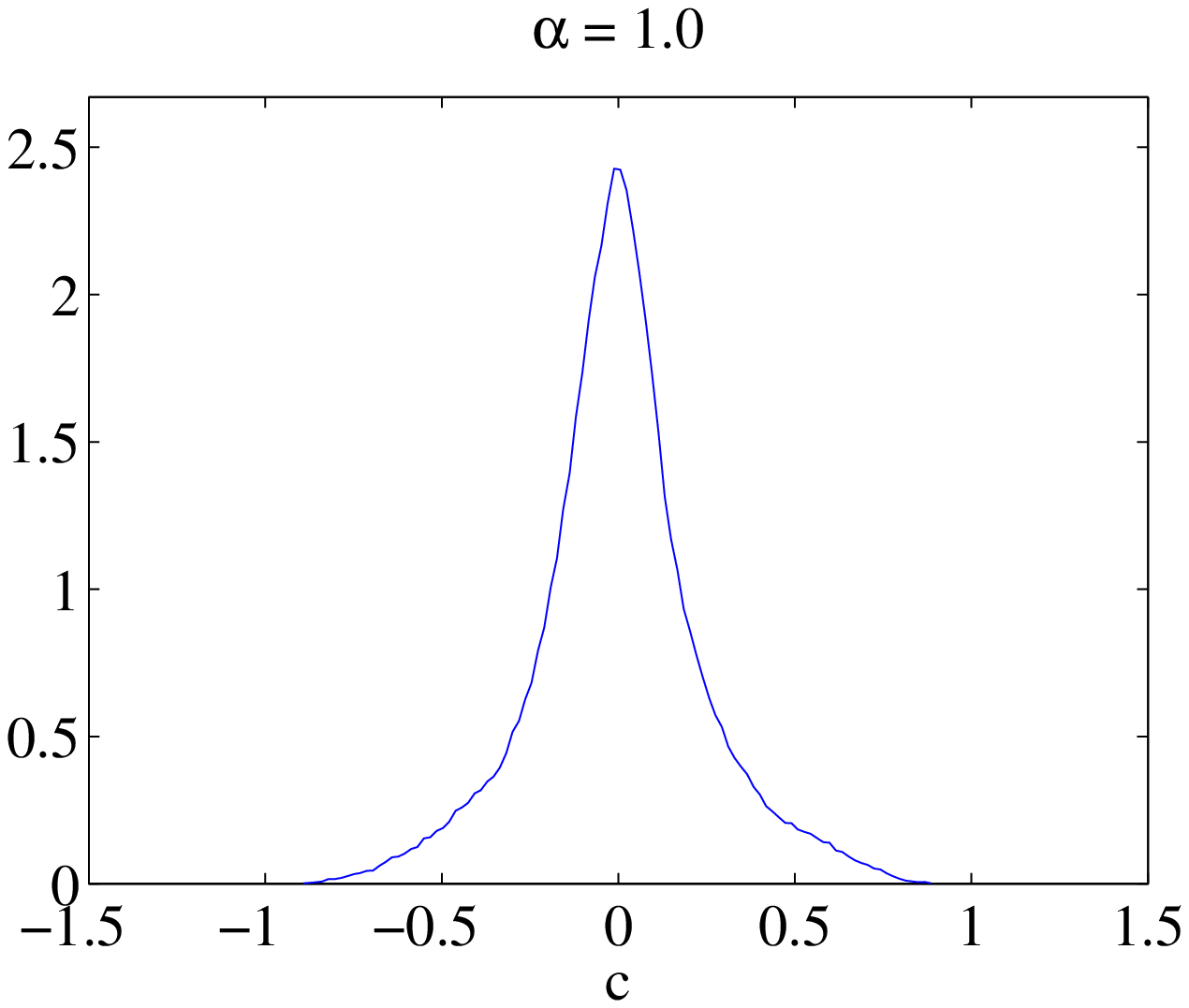}}}
\subfigure[]
{
     \scalebox{0.5}[0.5]{\includegraphics*[viewport=0 0 450 330]{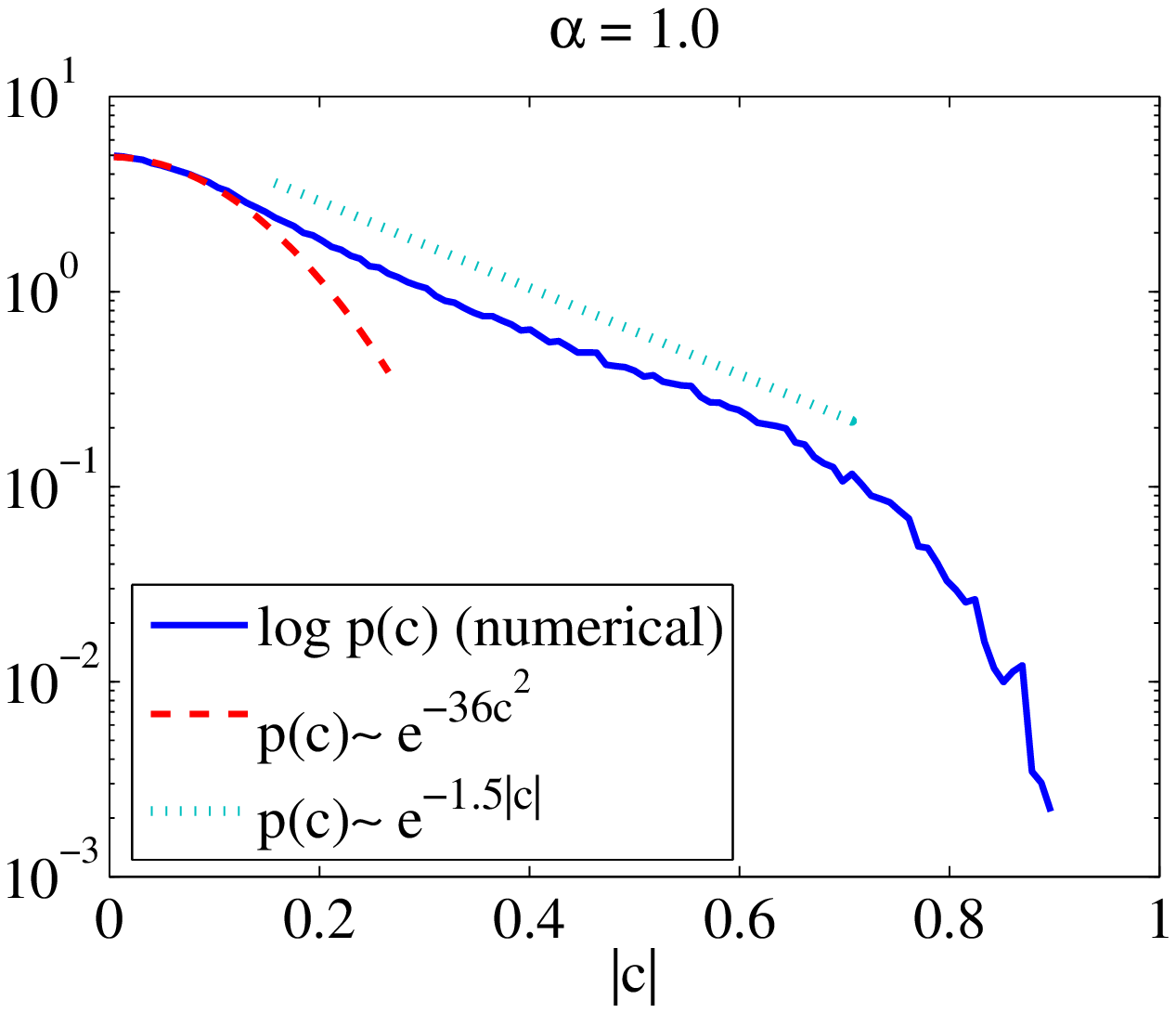}}}
\caption{%
Normalized PDF of concentration in the steady state. 
(a), (b) Segregation-dominated flow ($\alpha = 0.1, 0.3$); %
(c), (d) Crossover to quasi-diffusive regime ($\alpha = 0.5, 0.7$); %
(e) Quasi-diffusive regime ($\alpha = 1$); %
(f) Semilog plot of PDF for $\alpha=1.0$.  Note Gaussian core and exponential
 decay down to the cutoff at $\left|c\right|\lesssim1$.}
\label{fig:pdfs}
\end{figure}

In Fig.~\ref{fig:CHscale} the breakdown in the power law relationship $\overline{\sigma^2/F}\sim\lambda^{-1/3}$
for large stirring amplitudes suggests the crossover between a bubbly regime
and a diffusive one.  Another way of seeing the transition between these
regimes is to study the stationary probability distribution function (PDF)
of the field $c\left(\bm{x},t\right)$.  We show this in Fig.~\ref{fig:pdfs}.  For small
%
%
values of $\alpha$ we note that the PDF has sharp peaks at $\pm1$, indicating
the effectiveness of phase separation at these stirring amplitudes.  For
$\alpha = 1$ however, the PDF has a Gaussian core, indicating that
genuine mixing by advection-hyperdiffusion is taking place.  For intermediate
values of $\alpha$ the PDF is a combination of these two different distributions
of concentration.
\section{A One-Dimensional Equilibrium Problem}
\label{sec:1D}
Having observed the crossover between the bubbly and the well-mixed regimes,
we study a one-dimensional model to shed further light on the process
by which this occurs.

We examine the archetypal hyperbolic flow $\bm{v}=\left(-\lambda x,
\lambda y\right)$, with strain rate $\lambda$.  This flow tends to
homogenize the concentration in the $y$-direction through stretching. At
late times the problem then becomes one-dimensional,
\begin{equation}
  \frac{\partial c}{\partial t}-\lambda x\frac{\partial c}{\partial x}
  = D\frac{\partial^2}{\partial x^2}\left(c^3-c\right)-\gamma
  D\frac{\partial^4 c}{\partial x^4}.
\label{eq:C_H1D}
\end{equation}
Scaling lengths by $\sqrt{\gamma}$ and restricting to the steady case,
Eq.~\eqref{eq:C_H1D} becomes
\begin{equation}
  -\lambda\left(\frac{\gamma}{D}\right) x\frac{d c}{d x}
  = \frac{d^2}{d x^2}\left(c^3-c\right)-\frac{d^4 c}{d x^4}.
  \label{eq:CH1Dsteady}
\end{equation}
This equation is invariant under the parity change $x\rightarrow-x$ and so
we seek odd and even solutions.  Thus we may restrict the solution to the
half-line $\left[0,\infty\right)$.  We choose the bubble boundary conditions
$c\left(0\right)=1-\delta$, $c'\left(0\right)=0$, $c\left(\infty\right)=-1$,
$c'\left(\infty\right)=0$.  The addition of the small positive
constant $\delta$ in the boundary conditions makes a bubble solution possible
in the limit $\lambda\rightarrow0$.  In this limit, Eq.~\eqref{eq:CH1Dsteady}
reduces to
\begin{equation}
\frac{d^2}{dx^2}\left(c^3-c-\frac{d^2c}{dx^2}\right)=0
\label{eq:CH_bubble}
\end{equation}
and using the boundary conditions, this simplifies further to
\[
\tfrac{1}{2}\left(\frac{dc}{dx}\right)^2=\tfrac{1}{4}\left(c^2-1\right)^2+\left(-\tfrac{1}{2}\delta^2+\tfrac{1}{4}\delta^3\right)\left(c+1\right)\equiv\Phi\left(c\right),
\]
where the function $\Phi\left(c\right)$ is positive definite on the interval
$\left(-1,1-\delta\right)$.  This equation has the implicit solution
\[
x=\pm\frac{1}{\sqrt{2}}\int_{c\left(x\right)}^{1-\delta}\frac{dc}{\sqrt{\Phi\left(c\right)}}.
\]
The bubble width is given by the unique zero of the function $c\left(x\right)$.

\begin{figure}[htb]
\subfigure[]
{
   \scalebox{0.50}[0.50]{\includegraphics*[viewport=0 0 400 330]{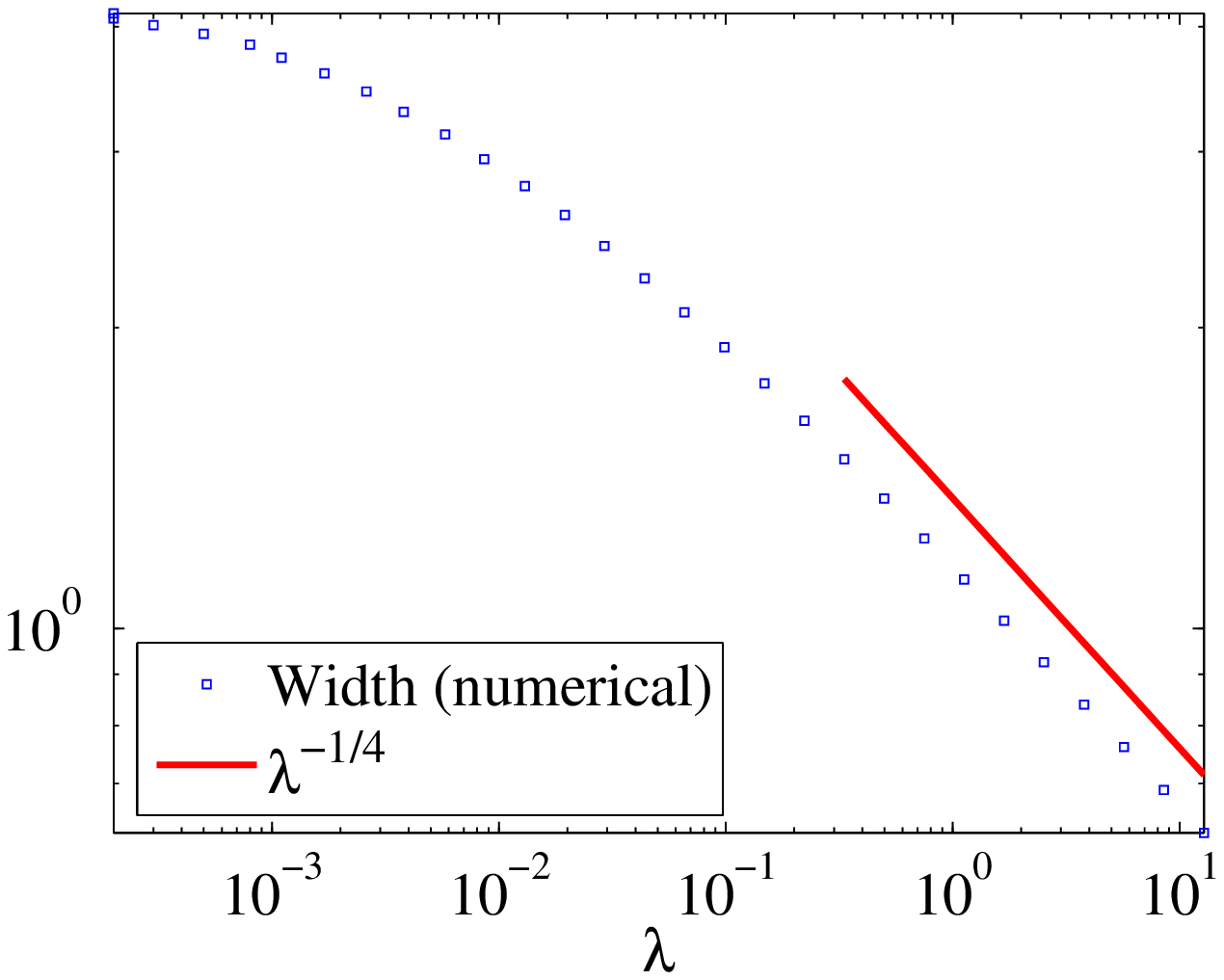}}
}
\hspace{0.01cm}
\subfigure[]
{
   \scalebox{0.50}[0.50]{\includegraphics*[viewport=0 0 400 330]{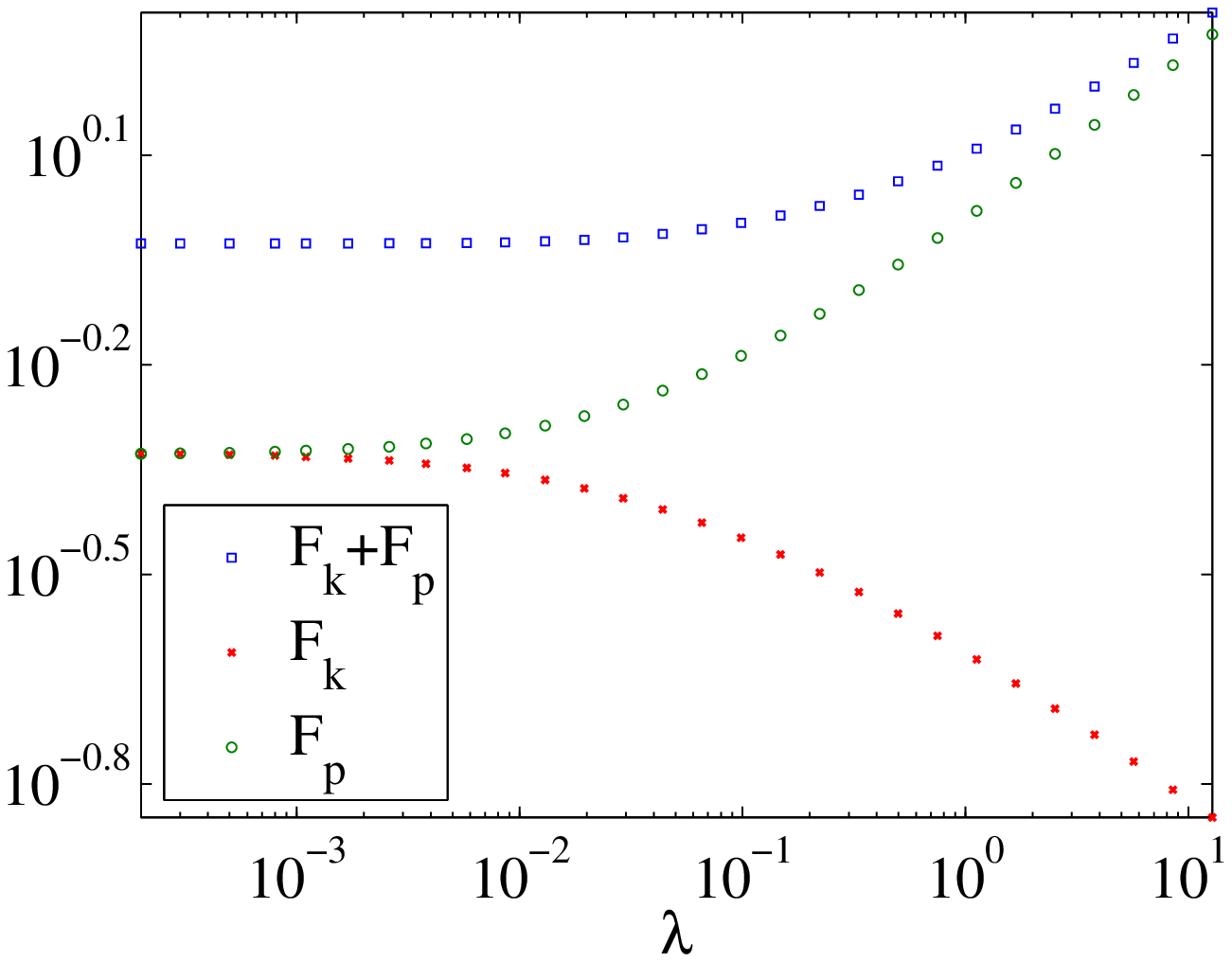}}
}
\subfigure[]
{
   \scalebox{0.50}[0.50]{\includegraphics*[viewport=0 0 400 330]{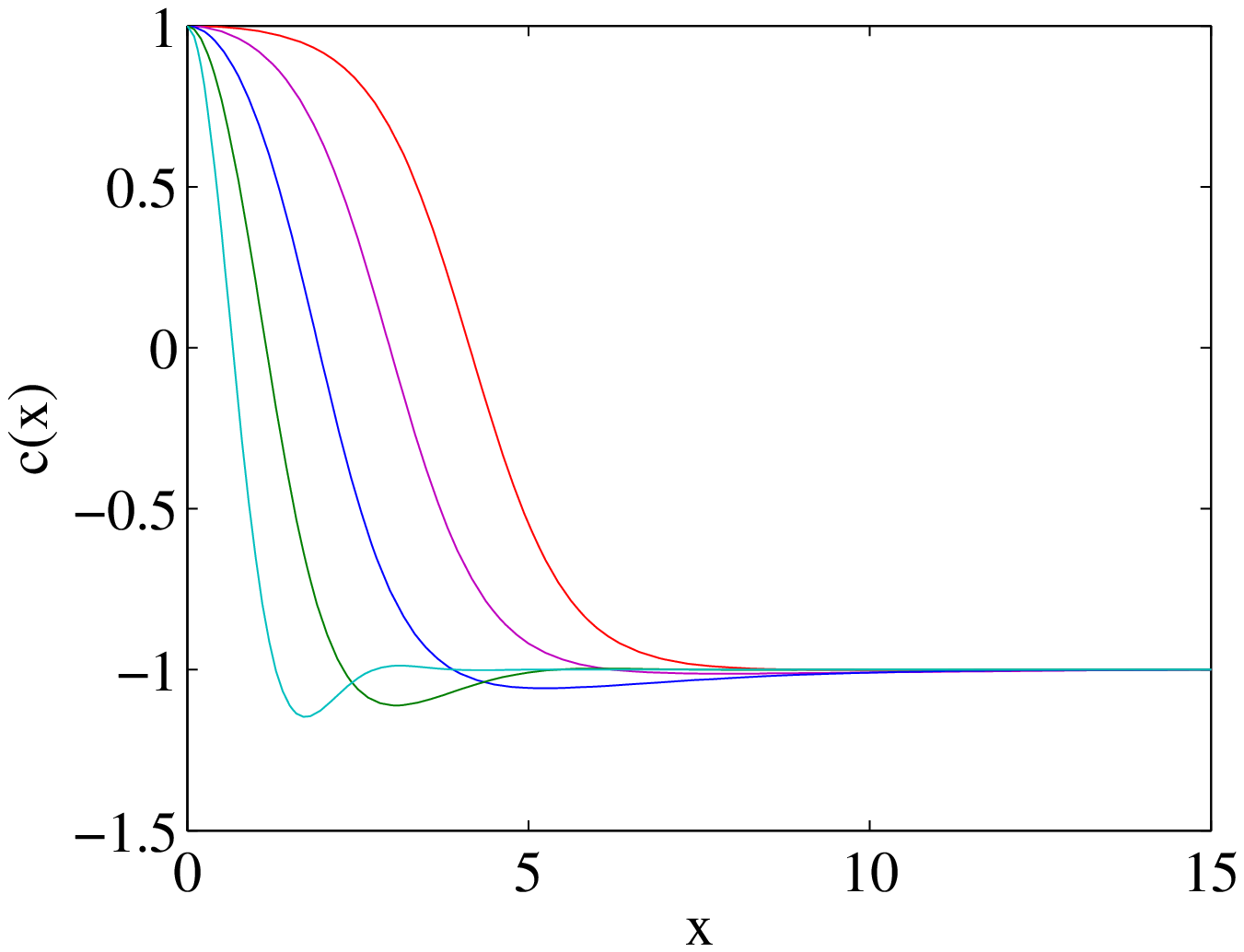}}
}
\caption{(a) Bubble width as a function of the strain rate $\lambda$; (b)
Bubble free energy as a function of the strain rate; (c) Bubble profile for
(from right to left) $\lambda = 0.001,0.01,0.1,1,10$.}
\label{fig:1D}
\end{figure}

By measuring lengths in terms of the diffusion lengthscale  $\left(\gamma
D/\lambda\right)^{1/4}$, the large-$\lambda$ limit of Eq.~\eqref{eq:CH1Dsteady} reduces to the advection-hyperdiffusion equation
\begin{equation}
x\frac{dc}{dx}=\frac{d^4c}{dx^4}.
\label{eq:hyperdiff}
\end{equation}
Restoring the dimensional units, the general solution is given in terms of
generalized Airy functions~\cite{Polyanin_odes},
\begin{equation}
c\left(x\right)=c\left(0\right)+\int_0^x dx'\sum_{\nu=0}^3 C_{\nu}v_{\nu}\left(x'\right);
\qquad v_{\nu}\left(x\right)=\epsilon_{\nu}\int_0^{\infty}\exp\left[\epsilon_{\nu}tx-\frac{t^4}{{4\lambda}/{\gamma
D}}\right]dt,
\label{eq:Airy_mod}
\end{equation}
where the constants $C_{\nu}$ sum to zero and the phases $\epsilon_{\nu}=\exp\left(i\pi\nu/2\right)$
 are the fourth roots of unity.  By inspection of the phases, we see that
 Eq.~\eqref{eq:hyperdiff} lacks a bubble solution, although
 it does have a front solution.  Since it is the segregation term which makes
 the existence of stationary bubbles possible, it is not surprising that
 no stationary bubble exists when this term is set to zero.  Thus as $\lambda\rightarrow\infty$
 the solution presented in Fig.~\ref{fig:1D} tends asymptotically to the
 function
\[
\left(1-\delta\right)+\int_{0}^x \left[ v_1\left(x'\right)-\left(1-i\right)v_2\left(x'\right)-iv_3\left(x'\right) \right]dx',
\]
which diverges for $x\rightarrow-\infty$.  Nevertheless, this solution on
the restricted interval $\left[0,\infty\right)$ provides the correct parametric
dependence of the length scale in the diffusive limit which suggests that
we are justified in using it.

We mention briefly an alternative set of boundary conditions: let $c\left(\pm\infty\right)=\pm
1$ and $c'\left(\pm\infty\right)=0$ (the $\delta$ term is unimportant
now).  This corresponds to a front.  At small $\lambda$, this equation contains
information about the width of the transition regions.  Since in this limit,
the front solution is independent of $\lambda$ and given by Eq.~\eqref{eq:CH_bubble},
there is no new information here.  At large $\lambda$ this solution has the
interpretation of being the transition layer between filaments, the width
of the layer giving the width of the filament.  In this limit, the solution
is given by $\int_0^{x}\left[v_1\left(x'\right)-v_3\left(x'\right)\right]dx'$.
 Thus, the front boundary conditions do not give any further information
 about the bubble / filament structures.
%
%

In Fig.~\ref{fig:1D} we provide the relation between the strain rate $\lambda$
and the bubble width and the bubble free energy.  For small $\lambda$ these
quantities depend only weakly on the strain rate while for large $\lambda$
the bubble width scales as $\lambda^{-1/4}$, in accordance with equation~\eqref{eq:R_diff}.
It is not clear if the free energy also has this asymptotic dependence. 
For large $\lambda$, the free energy equipartition rule is broken, and the
potential part dominates, in agreement with the results of Section~\ref{sec:numerics}.

Moreover, Fig.~\ref{fig:1D} indicates that the bubble width decreases
as the strain rate increases.  Thus, the length scale over which $c$ is constant
diminishes with increasing $\lambda$ so that for sufficiently large  $\lambda$,
the bubble size is less than or of the order of the transition region size.
 That is, $\left({\gamma D}/\lambda_{\mathrm{c}}\right)^{1/4}=\gamma^{1/2}$,
 so $\lambda_{\mathrm{c}}=D/\gamma$.
 On these scales, only diffusion is important and hence the tendency to form
 bubbles is overcome.  Thus, the one-dimensional model explains at least
 qualitatively why the crossover from the bubbly to the diffusive
 regime occurs.  This prediction for $\lambda_c$ is in agreement with the
 prediction in Section~\ref{sec:model} and is in approximate agreement with
 the numerical simulations, where the crossover occurred at $\alpha\simeq
 1$ ($\lambda\simeq 0.2$, $\tau D=\gamma$).

Finally we note that the concentration profile takes values
$\left|c\right|>1$ in some places, a consequence of Eq.~\eqref{eq:CH1Dsteady}
lacking a maximum principle, in contrast (say) to the advection-diffusion
equation.  This is consistent with the numerical results of Section~\ref{sec:1D},
where we find $\mathrm{max}_{\bm{x}\in V}\left|c\left(\bm{x},t\right)\right|>1$.
As mentioned by Gajewski and Zacharias \cite{Gajewski_nonlocal}, this
super-abundance of one binary fluid element is unreasonable on physical
grounds, pointing to a shortcoming in the advective CH model.  However, we
shall see in the next section that the qualitative features of this simple
model and the more realistic variable-mobility model are in agreement, suggesting
that the simpler model gives an adequate description of advective phase-separation
dynamics.
\section{The Variable-Mobility Case}
\label{sec:varmob}
In order to gain further insight into the dependence of the bubble size on
the Lifshitz-Slyozov and Lyapunov exponents, we study the variable-mobility Cahn--Hilliard equation \cite{Langer}
\begin{equation}\frac{\partial c}{\partial t} +\bm{v}\cdot\nabla c= \nabla\cdot\left[D\left(c\right)\nabla\mu\right],\end{equation}
where as before $\mu = f_0'\left(c\right)-\gamma\nabla^2c$.  By introducing
this additional feature, we recover a system that, at least in the case
without flow, has a solution $c\left(\bm{x},t\right)$ confined to the
range $[-1,1]$ \cite{Elliott_varmob}.  

We specify the functional form of the mobility as 

\begin{equation}D\left(c\right) = D_0\left(1-\eta c^2\right),\end{equation}
fixing $\eta = 1$.  This modification corresponds to interface-driven coarsening
because inside bubbles ($c=\pm 1$) the mobility is zero.  For other values
of $\eta$ we have a mixture of bulk- and interface-driven coarsening.  This
equation has been investigated by Bray and Emmott \cite{Bray_LSW} for $\bm{v}=0$,
in an evaporation-condensation picture.  They show that the typical bubble
size $R_{\mathrm{b}}\left(t\right)$ grows as $t^{1/4}$, a result limited
to dimensions greater than two.  Numerical simulations \cite{Zhu_numerics}
suggest that this growth law also holds in two dimensions.  We couple the
modified segregation dynamics to the sine flow.

For $\bm{v}=0$, we obtain the growth law for the scale $R_{\mathrm{b}}\equiv
1/k_1$, namely $R_{\mathrm{b}}\sim t^{0.264}$.  The growth exponent is close
to the value $1/4$ predicted by the LS theory.  In addition, we obtain the
free energy decay laws $F\sim t^{-0.267}$ again close to the LS value.  Thus,
 as in the constant mobility case, a description of the length scales in
 the system can be given in terms of the bubble energy or in terms of $R_{\mathrm{b}}$,
 with identical results.

Because the variable mobility calculation is computationally slower, we perform
the numerical experiments with flow at lower resolution.  We do not
anticipate that this will change the results, because in integrations of
the constant-mobility equations at resolution $256^2$ the scaling exponents
were unaffected.  Switching on the chaotic flow, we observe a steady state
characterized by fluctuation of the free energy, wave number $k_1$,
and variance $\sigma^2$ around mean values.  As in the constant mobility
case, the time-averaged free energy and time-averaged Batchelor scale possess
clear scaling laws,
%
%
\begin{figure}[htb]
\subfigure[]
{
   \scalebox{0.50}[0.50]{\includegraphics*[viewport=0 0 400 330]{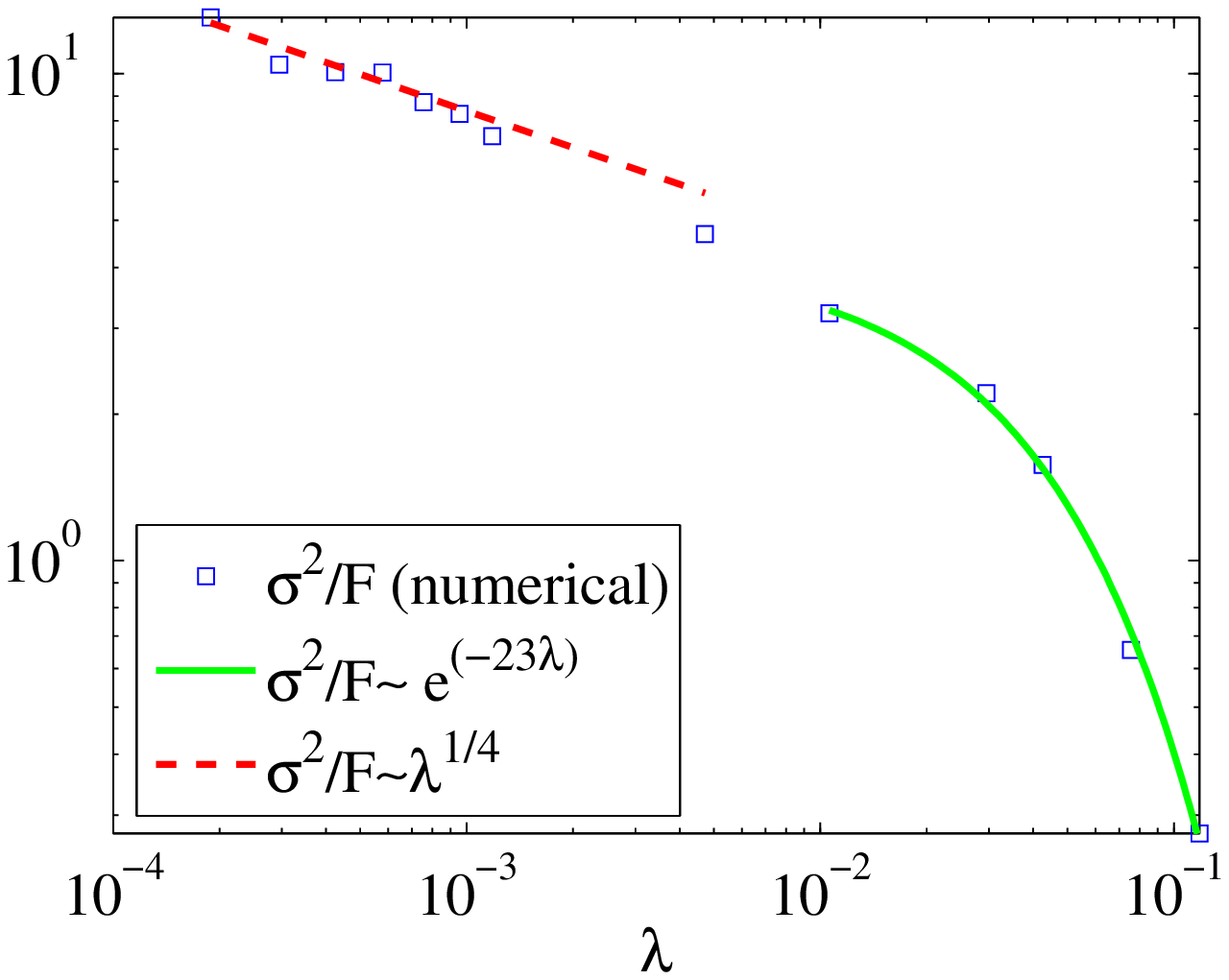}}
}
\hspace{0.01cm}
\subfigure[]
{
   \scalebox{0.50}[0.50]{\includegraphics*[viewport=0 0 400 330]{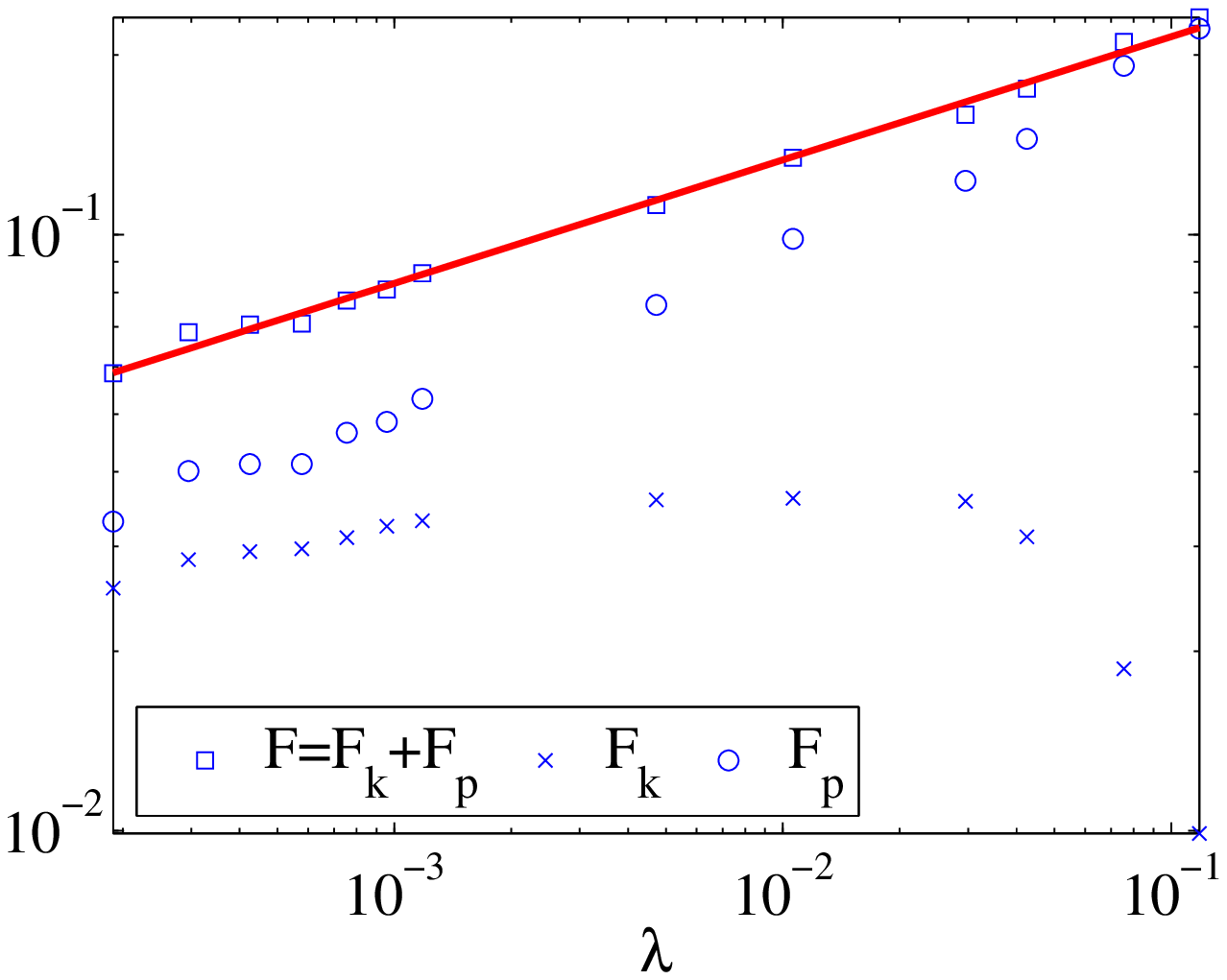}}
}
\caption{(a) 
The plot of $\log\big(\overline{\sigma^2/F}\big)$ against $\log\lambda$ suggests
the scaling law $\overline{\sigma^2/F}\sim\lambda^{-1/4}$ for small stirring
amplitudes, while for large amplitudes this proxy for the bubble radius decays
exponentially with increasing Lyapunov exponent.
(b) The graph of $\log\overline{F}$ exhibits power law behaviour $\overline{F}\sim\lambda^{0.21}$.
 The predominance of kinetic free energy over potential free energy at large
 $\lambda$ is visible.
}
\label{fig:varmob}
\end{figure}
while tthe mean wavenumber $\overline{k_1}$ does not.  For small
$\lambda$, the proxy for bubble radius $\overline{\sigma^2/F}$ scales
approximately as $\lambda^{-1/4}$, suggesting a decay of bubble radius
according to the LS exponent.  Thus the analogous result of
Section~\ref{sec:numerics} is not fortuitous.  For large $\lambda$ this
quantity decays exponentially to zero at a rate (with respect to $\lambda$)
close to that of the constant mobility system.  In each case, the fit of
the data is not as good as the constant mobility case, the computational
error being larger here because we have run the simulations at resolution
$256^2$.  We present these results in Fig.~\ref{fig:varmob}.

The results are an exact replica of those for the constant mobility case.
Thus, the arrest of bubble growth for small $\lambda$ is a genuine phenomenon
whose behavior depends on the LS exponent of the corresponding unstirred
dynamics.  The mixing at large $\lambda$ is identical to that for which the
mobility is constant.  Thus, while the variable mobility model is more
realistic than its constant-mobility counterpart, their properties with regard
to mixing are the same.

\section{Conclusions}
We have shown how, in the presence of an external chaotic stirring, the
coarsening dynamics of the Cahn--Hilliard equation are arrested and bubbles
form on a particular length scale.  The system reaches a steady state,
characterized by the fluctuation of the free energy $F$, the variance
$\sigma^2$, and the wavenumber $k_1$ around mean values.  A measure of the
typical bubble size is given by the time-average of $\sigma^2/F$.  For
sufficiently large stirring intensities, the bubbles diminish in prominence,
to be replaced by filament structures.  These are due to the combination
of the advection and hyperdiffusion terms in the CH demixing mechanism. 
In this regime, the concentration tends to homogenize, so that the stirring
stabilizes the previously unstable mixed state.

We have explained with a one-dimensional model how this transition arises
and how the filament width in the homogeneous regime depends on the Lyapunov
exponent of the chaotic flow.  This latter investigation exhibits the super-abundance
of one binary fluid element at a particular location in space, which we reject
as unreasonable.  In this result, we see how the lack of a
maximum principle for the constant-mobility CH equation limits its physical
relevance.  In the case of no flow this difficulty was overcome by the
addition of a variable mobility~\cite{Elliott_varmob}.  However, the
quasi-diffusive regime identified in our simulations is robust in the sense
that it is present in both the constant and the more realistic variable
mobility cases.  We therefore expect to see this remixing phenomenon in real
binary fluids.


\pagebreak

\end{document}